%% file: main.tex
\documentclass[10pt,journal,compsoc]{IEEEtran}
\input{header}

\ifCLASSOPTIONcompsoc
  \usepackage[nocompress]{cite}
\else
  \usepackage{cite}
\fi
\ifCLASSINFOpdf
\else
\fi
\begin{document}
%
\title{EpiTESTER: Testing Autonomous Vehicles with Epigenetic Algorithm and Attention Mechanism}
%
%
%
%

\author{Chengjie~Lu,
        Shaukat~Ali,
        and~Tao~Yue 
\IEEEcompsocitemizethanks{\IEEEcompsocthanksitem C. Lu is with the Department
of Engineering Complex Software Systems, Simula Research Laboratory, Oslo, 0258, Norway. 
E-mail: chengjielu@simula.no.
\IEEEcompsocthanksitem S. Ali is with Simula Research Laboratory and Oslo Metropolitan University, Oslo, 0258, Norway. 
E-mail: shaukat@simula.no.
\IEEEcompsocthanksitem T. Yue (the corresponding author) is with the School of Computer Science and Engineering, Beihang University, Beijing, 100191, China. 
E-mail: yuetao@buaa.edu.cn.
}
}

\IEEEtitleabstractindextext{%
\begin{abstract}
Testing autonomous vehicles (AVs) under various environmental scenarios that lead the vehicles to unsafe situations is known to be challenging. Given the infinite possible environmental scenarios, it is essential to find critical scenarios efficiently. To this end, we propose a novel testing method, named \method, by taking inspiration from epigenetics, which enables species to adapt to sudden environmental changes. 
In particular, \method adopts gene silencing as its epigenetic mechanism, which regulates gene expression to prevent the expression of a certain gene, and the probability of gene expression is dynamically computed as the environment changes. Given different data modalities (e.g., images, lidar point clouds) in the context of AV, \method benefits from a multi-model fusion transformer to extract high-level feature representations from environmental factors and then calculates probabilities based on these features with the attention mechanism. To assess the cost-effectiveness of \method, we compare it with a classical genetic algorithm (GA) (i.e., without any epigenetic mechanism implemented) and \method with equal probability for each gene. We evaluate \method with four initial environments from CARLA, an open-source simulator for autonomous driving research, and an end-to-end AV controller, Interfuser. Our results show that \method achieved a promising performance in identifying critical scenarios compared to the baselines, showing that applying epigenetic mechanisms is a good option for solving practical problems.
\end{abstract}

}

\maketitle

\IEEEdisplaynontitleabstractindextext

%
\IEEEpeerreviewmaketitle

\IEEEraisesectionheading{\section{Introduction}\label{sec:introduction}}
\IEEEPARstart{S}{imulation}-based testing has become a widely applied method for testing autonomous vehicles (AVs)~\cite{tuncali2018simulation,10.1109/ICSE48619.2023.00155}. Such testing typically requires simulating environmental scenarios characterized by many parameters, such as weather conditions and pedestrian behaviors. The possible combinations of such parameter configurations could potentially be infinite. Therefore, cost-effectively searching for environmental scenarios with a high chance of leading an AV to collisions and other safety violations is an optimization problem.
Existing testing methods typically employ search-based optimization to select a subset of environmental parameters and treat these parameters equally when generating AV test scenarios. However, not all parameters contribute equally in a given driving status~\cite{wang2019exploring,van2020path}; for example, road users such as pedestrians contribute more to the complexity of urban driving than weather parameters, as urban roads have more complex road features such as signalized intersections~\cite{wang2019exploring}, while in highway testing, weather parameters can be more critical because vehicle speeds are usually high and adverse weather conditions may significantly affect braking distances~\cite{kordani2018effect}. Thus, a search-based optimization could benefit from selectively disabling exploration and exploitation of specific parameters (i.e., \textit{silencing} them in evolution) such that the complexity of the optimization problem can be reduced and faster convergence to optimal solutions can be potentially achieved. 

To achieve this goal, 
epigenetics~\cite{weinhold2006epigenetics,gibney2010epigenetics}, which studies how genes are regulated and expressed without altering the DNA code, is an optional and innovative solution. It provides insights into how gene expression is regulated and how various factors, such as environmental exposures, can influence it. 
In biology, various epigenetic mechanisms (EMs) have been studied, such as histone modifications, imprinting, and gene silencing (GS), which regulate gene expression, development, and adaptation to environmental changes. 
In this paper, we study whether the GS mechanism from epigenetics can be employed in a genetic algorithm (GA) to regulate gene (parameter) expression during search, i.e., prioritizing environmental parameters with a high probability of leading to collisions or other safety violations, with the ultimate goal of improving search efficiency in finding critical environmental configurations (i.e., driving scenarios) for AV testing.

In the literature, a set of solutions (e.g., based on search algorithm~\cite{ben2016testing,abdessalem2018testing,9251068}, reinforcement learning~\cite{lu2022learning,10.1109/ICSE48619.2023.00155,feng2023dense}, and real-world traffic reports~\cite{zhang_risk_2022,feng2023dense,yan2023learning}) have been proposed for identifying and generating critical driving scenarios. Some formulate the optimization problem into a search problem and solve it with well-known search algorithms such as GAs~\cite{holland1992genetic} and Non-dominated Sorting Genetic Algorithms (NSGA)~\cite{deb2002fast}. However, none of them applies epigenetic algorithms (e.g., epiGA~\cite{STOLFI2018250}), which, in our opinion, suits well for simulation-based testing of AVs, mainly because GA and NSGA pass down genes from parents to their offspring via genetic inheritance but \textit{epigenetic inheritance} allows for \textit{fast adaptation when appropriate} via regulating how genes work, which potentially speeds up convergence while keeping stability in the ever-changing operating environment of AVs, as discussed in~\cite{tanev_epigenetic_2008,VERSE2023}.

To this end, we propose a novel approach called \method, which mimics the \textit{GS} mechanism -- silencing one or more genes (i.e., environmental parameters in testing AVs) with GS probabilities (i.e., probabilities of silencing environmental parameters);
thereby focusing on parameters that are highly likely to contribute to leading an AV to an unsafe situation.
To dynamically generate probabilities for GS as the environment changes, \method first uses a multi-model fusion transformer to extract environmental features from various data modalities (camera images, LiDAR) and then passes the extracted features to a self-attention layer to predict GS probabilities.

In the literature, several epigenetic algorithms (e.g., epiGA~\cite{STOLFI2018250}, EpiLearn~\cite{mukhlish2020reward} and RELEpi~\cite{mukhlish2020reward2}) have been proposed and applied for solving benchmark problems such as the multidimensional knapsack problem~\cite{kellerer2004multidimensional}. To the best of our knowledge, \method is the very first work of encoding and solving the problem of searching for critical driving scenarios with epigenetic algorithms. Though \method is based on epiGA, it implements a novel GS mechanism and a completely new epigenetic model trained for generating GS probabilities dynamically as the environment state changes, i.e., in each simulation cycle.

We evaluated \method with a state-of-the-art AV controller (i.e., Interfuser~\cite{shao2023safety}) and a commonly used simulator (i.e., CARLA~\cite{dosovitskiy2017carla}). We compared \method with two baseline methods: a classical GA and a modified \method with equal GS probabilities, i.e., \methodequal. The evaluation results show that \method outperformed the baselines, and the GS mechanism based on the attention mechanism can effectively differentiate the contribution of each environmental parameter to safety violations, including collisions, and express parameters with higher contributions.

In summary, our contributions are: 1) a novel formulation of the problem of searching for critical driving scenarios with epigenetics, 2) an epigenetic model based on a multi-model fusion transformer and attention mechanism to predict probabilities for gene silencing, 3) a novel method, i.e., \method, integrating the epigenetic algorithm and the attention mechanism, 
and 4) an empirical evaluation demonstrating the benefits of \method over the baselines. 

The rest of the paper is organized as follows. Section~\ref{sec:background} introduces the background, including the epigenetic algorithm, attention mechanism, and transformer model. We present our optimization problem formulation in Section~\ref{sec:problem_representation} and introduce \method approach in Section~\ref{sec:method}. We then introduce the experiment design in Section~\ref{sec:experiment_design} and report the evaluation results in Section~\ref{sec:results}, which is followed by discussions in Section~\ref{sec:discussions}. Finally, we report the related work in Section~\ref{sec:related_work} and conclude the paper in Section~\ref{sec:conclusion}.

\section{Background}\label{sec:background}

\subsection{Epigenetic Algorithm}\label{subsec:epiga}
Epigenetics is the study of changes in organisms caused by gene expression modifications, which are affected by various factors, such as individual behaviors and environmental exposures.
Essentially, epigenetic changes are heritable changes that don't involve modifications to DNA sequences but can introduce uncertainties into gene expressions, consequently boosting species' chances of survival~\cite{bollati2010environmental}. One interesting example is that octopuses change their color, shape, and texture in real-time in response to environmental stimuli, which is not pre-determined by their genes~\cite{liscovitch2017trade}. Several recent reviews have studied the role of epigenetics in domesticated animals~\cite{vogt2017facilitation}, plants~\cite{lamke2017epigenetic}, and humans~\cite{kanherkar2014epigenetics}. 
Epigenetics studies reveal that genes passed down from parents to offspring via genetic inheritance cannot deal with sudden environmental changes. However, via \textit{epigenetic inheritance}, one can achieve \textit{fast adaptation when appropriate} by controlling genes (e.g., turning them on or off).

In particular, epiGA~\cite{STOLFI2018250} implements the GS mechanism and integrates it into GA to control how genes are expressed in response to environmental changes. In epiGA, an individual is composed of multiple cells, each consisting of a chromosome and a nucleosome. The chromosome uses multiple genes (i.e., parameters) to encode a solution to the problem, and the nucleosome is a binary mask of the same length as the chromosome that controls the accessibility of genes. Specifically, a position equal to 1 in the nucleosome indicates that the corresponding gene is collapsed (i.e., the gene is inaccessible and cannot be changed during reproduction), while a 0 in the nucleosome indicates that the corresponding gene is uncollapsed, that is, the gene is accessible during reproduction.
The GS mechanism applied in epiGA controls the gene expression through DNA methylation~\cite{bender2004dna}, one of the major epigenetic modifications controlling gene expression. In GS, only collapsed genes have a chance to be methylated, and the methylation probability of each gene (hereinafter referred to as GS probability) is provided by the environment.


\subsection{Attention Mechanism and Transformer}\label{subsec:attention}
The attention mechanism is crucial in modern machine learning applications, such as natural language processing~\cite{bahdanau2016neural} and computer vision~\cite{guo2022attention}. It mimics cognitive attention by enabling machine learning models to focus selectively on specific parts of input data, enhance their ability to capture long-range dependencies and improve performance on complex tasks. In particular, self-attention~\cite{zhang2019self}, as a specific type of attention mechanism, allows a model to dynamically weigh the significance of individual elements within the same data sequence.

The Transformer network architecture introduced by Vaswani et al.~\cite{vaswani2017attention} processes sequential data, and its self-attention mechanism is often called scaled dot-product attention. In the Transformer architecture, each element in the input sequence is first embedded into three vectors: a query vector (Q), a key vector (K), and a value vector (V). Then, for each element in the sequence, the self-attention mechanism computes the attention outputs:
\begin{equation}
    \text{Attention(Q, K, V)} = softmax(\dfrac{Q\cdot K^T}{\sqrt{d_k}})\cdot V,
\end{equation}
where $\sqrt{d_k}$ is the scaling factor and the \textit{softmax} function is the default function of converting attention scores (${Q\cdot K^T}/{\sqrt{d_k}}$) into probabilities. In \method, we opt for \textit{sigmoid} because the prediction of GS probabilities is a multi-label problem (i.e., multiple non-mutually exclusive GS probabilities of selecting or silencing multiple parameters).


\section{Problem Representation}\label{sec:problem_representation}
In our context, the test environment is about a simulated AV under test (\avut) with an autonomous driving system (\ads) deployed drives in a virtual environment, where driving scenarios (also commonly called test scenarios in AV testing) characterized with various environmental parameters such as pedestrians, NPC vehicles, and weather conditions, are simulated. Below, we first describe a list of configurable environmental parameters that characterize test scenarios (Section~\ref{subsec:env_para}), then define the formulation of the optimization problem (Section~\ref{subsec:optimization_pro}).

\subsection{Configurable Environmental Parameters}\label{subsec:env_para}
AV's safety is affected by various environmental factors~\cite{mind_gap}. For example, the fog density affects the AV's perception module, which could degrade the AV's safety. Though there are infinite environmental factors in the real world, when it comes to a simulated environment, the number of parameters that can be simulated and configured is limited and subject to the capability of the employed simulator. Based on the simulator we use in this paper, we present two categories of configurable environmental parameters, discussed below.

\subsubsection{Dynamic object parameters}\label{subsec:obj_para}
These parameters characterize objects with dynamic behaviors, such as pedestrians and NPC vehicles. Including them in AV testing is crucial due to the inherent uncertainty and difficulty in predicting their behaviours~\cite{ridel2018literature,mozaffari2020deep}. 
In our current design of \method, we consider two types of dynamic objects: pedestrians and NPC vehicles.

\begin{compactitem}
    \item A pedestrian ($ped$) is characterized with a 5-tuple specifying its initial position, behavior, and speed: $<$\textit{$dis^{lo}_{ped}$}, \textit{$dis^{la}_{ped}$}, \textit{$o^x_{ped}$}, \textit{$o^y_{ped}$}, $v_{ped}$$>$, where $dis^{lo}_{ped}$ and $dis^{la}_{ped}$ are the distances of $ped$ from the \avut in the longitudinal and lateral directions respectively, and $o^x_{ped}$ and $o^y_{ped}$ denote $ped$'s orientation. Its initial speed is denoted as $v_{ped}$.
    \item An NPC vehicle ($npc$) is characterized with a 3-tuple denoting its initial position and behavior: $<$\textit{$dis^{lo}_{npc}$}, \textit{$dis^{la}_{npc}$}, $ behavior_{npc}$$>$. The initial position ($dis^{lo}_{npc}$ and $dis^{la}_{npc}$) is the distances of $npc$ from \avut in the longitudinal and lateral directions. Our goal is to generate realistic test scenarios; therefore, we consider specifying the initial behavior (i.e., $ behavior_{npc}$) of $npc$ and having its subsequent behaviors controlled by a control policy from the simulator. The policy navigates to a destination while avoiding potential collisions as much as possible.
\end{compactitem}

\textit{\textbf{Parameter Ranges}}. Ranges for the pedestrian and \npc parameters define valid inputs of test scenarios. Based on the results of our pilot study, we define:
\begin{inparaenum}[1)]
    \item the ranges for the initial position parameters of the pedestrian (i.e., $dis^{lo}_{ped}$ and $dis^{la}_{ped}$) as $[-10m, 10m]$, corresponding to 10 meters behind (left) of the \avut to 10 meters ahead (right) of the \avut; 
    \item the range for the orientation parameters of the pedestrian (i.e., $o^x_{ped}$ and $o^y_{ped}$) as $[-1, 1]$;
    \item the orientation of the pedestrian ($o^x_{ped}$, $o^y_{ped}$) as a vector on the coordinate system with the vector direction indicating the orientation of the pedestrian, and with the values from -1 to 1 determining the amount of rotation in the counterclockwise or clockwise direction along the x-axis/y-axis;
    \item the pedestrian's speed (i.e., $v_{ped}$) as the average human walking speed is between 0.94m/s and 1.43m/s, according to the National Institutes of Health~\cite{humanwalkspeed}; 
    \item the ranges of the initial position parameters of \npc (i.e., $dis^{lo}_{npc}$ and $dis^{la}_{npc}$) as $[-20m, 20m]$, corresponding to 20 meters behind (left) to 20 meters ahead (right) of the \avut; and 
    \item three possible initial \npc behaviors: maintaining the lane, changing to the right lane, and changing to the left lane, as evidence has shown they are typical behaviors involving adversarial interactions between the \avut and its surrounding \npc~\cite{lu2022learning,chen2021adversarial}.
\end{inparaenum}

Furthermore, according to Ro et al.~\cite{ro2020new}, a vehicle should keep a safety distance of at least 5 meters away from its surrounding objects to avoid potential safety violations; therefore, to ensure the realism of the test scenarios, i.e., the potential safety violations may happen during the driving process instead of the moment the environment is configured, we require that the initial distance between the \npc and the \avut should be no less than 5 meters.

\subsubsection{Weather parameters}\label{subsec:wea_para}
Weather conditions can greatly impact AV decision-making~\cite{8500659}. \method considers the sun altitude angle and the fog density, i.e., $<$\textit{$angle_{sun}$}, $density_{fog}$$>$, two essential weather parameters in AV testing~\cite{morl}. Configuring the sun altitude angle affects illumination conditions, e.g., shadows, direct sunlight, or over/underexposed, which can degrade the performance of vision-based modules of an AV~\cite{8691698}. The density of fog affects the visibility of the environment, thereby affecting the AV perception module.

\textit{\textbf{Parameter Ranges}}. We follow the weather parameter ranges of the simulator; the sun altitude angle is from -90 (midnight) to 90 (midday), and the fog density is from 0 (clear) to 100 (invisible).

\subsection{Problem Formulation}\label{subsec:optimization_pro}
Generating test scenarios with a high probability of revealing safety violations can be formulated as a search problem. The entire search space $SS$ is all possible combinations of values of the configurable environmental parameters. Note that values of all parameters are numeric, except for the behavior of \npc, which is categorical; therefore, the number of possible solutions (i.e., test scenarios) is infinite, and exhaustively exploring the entire search space is practically infeasible.

\textit{\textbf{Test Inputs and Outputs}}. A feasible test input is a vector with 10 values of all configurable environmental parameters, i.e., $<$\textit{$dis^{lo}_{ped}$}, \textit{$dis^{la}_{ped}$}, \textit{$o^x_{ped}$}, \textit{$o^y_{ped}$}, $v_{ped}$, \textit{$dis^{lo}_{npc}$}, \textit{$dis^{la}_{npc}$}, $ behavior_{npc}$, \textit{$angle_{sun}$}, $density_{fog}$$>$. A \textit{test input} configures the initial state of a test scenario, such as the initial position of a \npc, and the initial walking speed of a pedestrian.
Executing a test scenario with a test input generates a list of time-stamped outputs containing the test environment states and safety measurements, i.e., \textit{test outputs}, which are then used to compute a fitness value with the fitness function.
In our current design, we define the fitness function as the minimum distance between the \avut and other objects ($obj$):
\begin{equation}
    \mathcal{F}(\avut, obj)=\min\limits_{o\in obj}Dis_o(\avut, o),
\end{equation}
where, $Dis()$ is the Euclidean distance formula, which has been shown effective in estimating the distance between the \avut and obstacles in AV testing research~\cite{ben2016testing,lu2022learning}. Hence, our test scenario generation problem can be formulated as the following optimization problem.

\textit{\textbf{Optimization Problem}}. Given a set of environmental parameters and their value ranges, find an optimal solution $s^*$ from the search space $SS$ as the test input that satisfies:
\begin{equation}
    \forall s_i \in SS \cap s_i \neq s^*: \mathcal{F}_{s^*}(\avut, obj_{s^*}) \leq \mathcal{F}_{s_i}(\avut, obj_{s_i})
\end{equation}

\section{\method Approach}\label{sec:method}


\begin{figure*}[htbp]
  \centering
  \includegraphics[width=\textwidth]{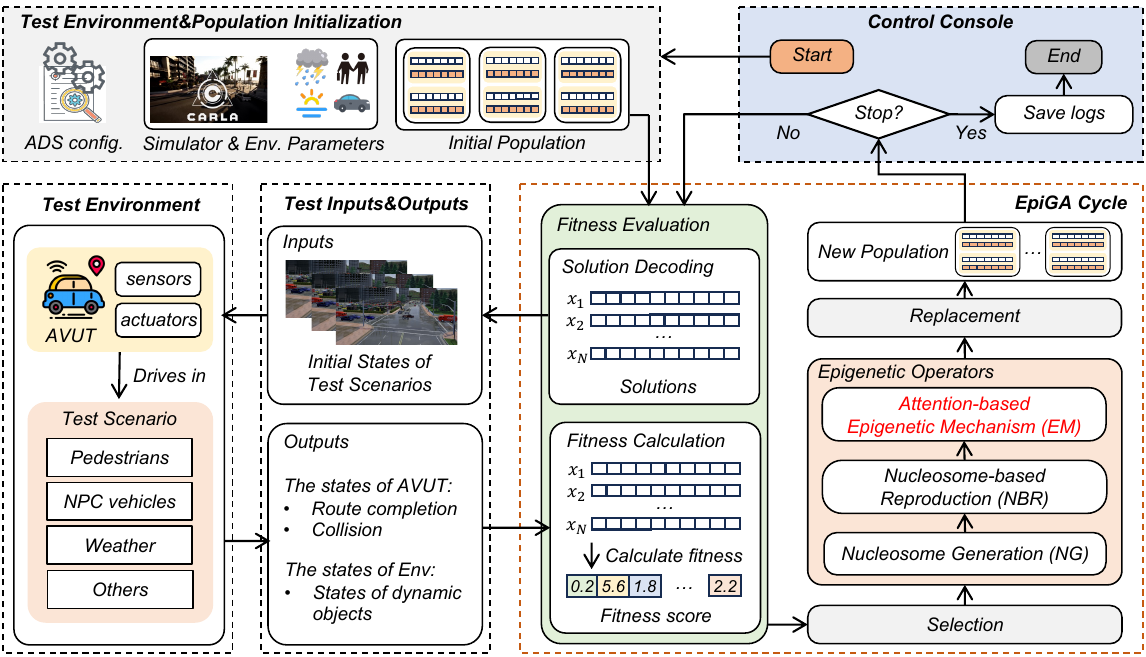}
  \caption{\textbf{Overview of \method}. \textit{Control Console} controls the start and end of the \method workflow; \textit{Test Environment$\&$Population Initialization} initializes the test environment and population for \epiga; \textit{Test Environment} is a simulated environment where the \avut drives in various test scenarios; \textit{Test inputs$\&$Outputs} specify the inputs and outputs for the AV testing problem that \method solves; \textit{EpiGA Cycle} is the place where \epiga is employed, and the \textit{Epigenetic Mechanism (EM)} has been extended based on the attention mechanism by \method.}
  \label{fig:overview}
\end{figure*}
Figure~\ref{fig:overview} depicts the overall working mechanism of \method. 
\textit{Control Console} starts the workflow and evaluates the stopping criteria in each cycle. Each cycle starts with \textit{Test Environment$\&$Population Initialization}. Specifically, the test environment is initialized, i.e., starting the simulator and loading the configurable environmental parameters, configuring the \ads, and deploying it to the \avut. The initialization generates an initial population containing all encoded initial test solutions. Each encoded solution needs to be decoded into a test input (i.e., the initial state of the test scenario) during \textit{Solution Decoding} before being fed into \textit{Test Environment}. \textit{Test Environment} then simulates test scenarios where the \avut drives in. After that, a list of outputs, including states of the \avut and the environment, is returned to \textit{Fitness Calculation} for each solution's fitness score. 
Then, \method's \textit{EpiGA Cycle} starts, which is essentially the application of a sequence of genetic and epigenetic operators: the \textit{Selection} operator, three \textit{Epigenetic Operators}, and the \textit{Replacement} operator. We apply GS as the \textit{Epigenetic Mechanism (EM)}, which controls the expression of each gene based on the attention mechanism.
More details about the \textit{EpiGA Cycle} will be introduced in Section~\ref{subsec:epiga_cycle},
and the attention-based GS probability generation will be introduced in Section~\ref{subsec:model}. Eventually, a new population is generated, and the workflow continues if the termination condition is not met; otherwise, the workflow terminates.

\subsection{EpiGA-guided Scenario Generation}\label{subsec:epiga_cycle}
We employ \epiga~\cite{STOLFI2018250} as our optimization algorithm. 
Since the optimization problem encoding is novel, we provide detailed discussions of each genetic and epigenetic operator in the subsequent subsections. Below, we first describe three genetic operators: initialization, selection, and replacement in Section~\ref{subsubsec:population_initialization}, then present two epigenetic operators (i.e., nucleosome generation and nucleosome-based reproduction) in Section~\ref{subsubsec:ng_nbr}, and finally, we introduce the epigenetic mechanism that \method applies, i.e., GS, in Section~\ref{subsubsec:gs}.

\subsubsection{Population Initialization, Selection and Replacement}\label{subsubsec:population_initialization}
\begin{figure*}[htbp]
  \centering
  \includegraphics[width=\linewidth]{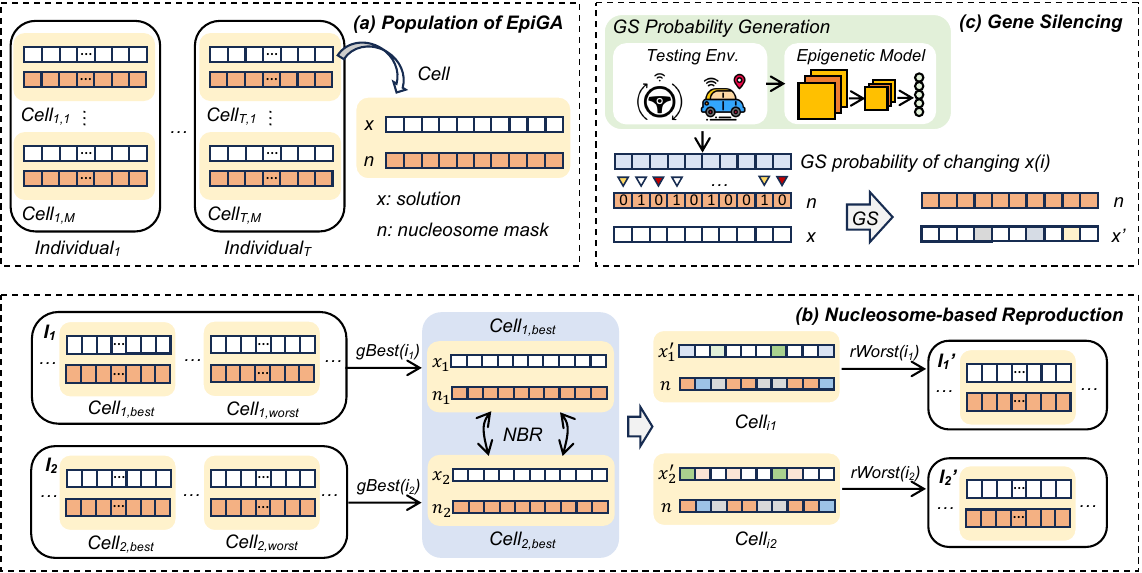}
  \caption{\textbf{\method's Population Structure, Nucleosome-based Reproduction, and Gene Silencing}. \textbf{Sub-figure (a)} illustrates that the population $P$ contains $T$ individuals each with $M$ cells. Each cell has two vectors: solution $x$ and nucleosome mask $n$; \textbf{sub-figure (b)} illustrates the nucleosome-based reproduction (NBR) process of reproducing new offspring by recombing solutions from parents with the guidance of nucleosome mask, among which functions $gBest()$ and $rWorst()$ obtain the best cell from an individual for \textit{NBR} and replace the worst cell with new offspring, respectively; \textbf{sub-figure (c)} illustrates the epigenetic mechanism that \method has extended and applied, i.e., gene silencing.}
  \label{fig:operators}
\end{figure*}

A population $P$ contains a set of $T$ individuals generated randomly. Once it is created, the EpiGA cycle begins. In contrast to the conventional behavior, where each individual encodes a solution to the problem being solved, in \epiga, each individual has $M$ cells. Each cell represents a distinct solution to the problem. As illustrated with the example shown in Figure~\ref{fig:operators}(a), each cell of an individual has two vectors (i.e., $x$ and $n$).
The length of each vector is equal to the number of configurable parameters (Section~\ref{subsec:optimization_pro}).
The $x$ vector encodes the solution, whereas
$n$ represents the nucleosome structure for the binary mask, which controls changes in each gene with a binary encoding during nucleosome-based reproduction. Concretely, the value in position $j$ of $n$ being 1 (i.e., $n_j=1$) means that the same position in the solution $x$ (i.e., $x_j$) is unchangeable (i.e., collapsed); if $n_j$ equals 0, it means that $x_j$ can be changed (i.e., uncollapsed).


We use binary tournament selection~\cite{GOLDBERG199169} as the selection operator, commonly applied in GAs. This operator selects the fittest individuals from the current generation and passes them to subsequent operators.

As for the replacement operator, we employed the elitist replacement~\cite{10.5555/534133} to select and preserve the best individuals for the next generation.

\subsubsection{Nucleosome Generation (NG) and Nucleosome-based Reproduction (NBR)}\label{subsubsec:ng_nbr}
For population $P$, \textit{NG} generates a new nucleosome vector ($n$) as a mask for each cell in each individual based on nucleosome probability $Pr_n$ and nucleosome radius $R$. In detail, for position $k$ of $n$ (i.e., $n_k$), if a randomized value $rand()$ is less than $Pr_n$, 
\textit{NG} starts to collapse positions around $n_k$, i.e., setting their values as 1. Which positions to be collapsed around $n_k$ are determined by $R$: $max(\textit{k-R}, 0)$ to $min(\textit{k+R}, len(n))$.

The \textit{NBR} operator reproduces new offspring by recombining solutions from parents, guided by the nucleosome mask. As shown in Figure~\ref{fig:operators}(b), \textit{NBR} performs on two individuals (i.e., $I_1$ and $I_2$) from the current population $P$. First, according to fitness values, the best cells $c_1$ and $c_2$ are extracted from $I_1$ and $I_2$, respectively. Next, \textit{NBR} is performed on $c_1$ and $c_2$ to produce new cells. As described in Algorithm~\ref{alg:nbr}, to generate new cells, \textit{NBR} first calculates a new nucleosome mask $n$ as $n_1 \textbf{OR} n_2$, where $n_1$ and $n_2$ are nucleosomes of $c_1$ and $c_2$.
The logic is that for a position $j$, if one of $n_1(j)$ and $n_2(j)$ is collapsed (i.e., 1), then the corresponding position in $n$ will also be collapsed.
Then, for those uncollapsed positions (i.e., 0), their corresponding values of $x_1$ and $x_2$ in those positions are swapped (Lines 6-12), and new solutions $x_1^\prime$ and $x_2^\prime$ are then produced. Eventually, two new cells $c_1$ and $c_2$ (Line 14) are created with the new nucleosome and solutions. The two new cells replace the two worst cells in $I_1$ and $I_2$; consequently, two new individuals are generated, i.e., $I_1^\prime$ and $I_2^\prime$.

\begin{algorithm}[htbp]
    \small
  \caption{Nucleosome-based Reproduction (NBR)}
  \label{alg:nbr}
  \input{algs/nbr}
\end{algorithm}

\begin{algorithm}[htbp]
    \small
  \caption{Gene Silencing (GS)}
  \label{alg:gs}
  \input{algs/gs}
\end{algorithm}

\subsubsection{Epigenetic Mechanisms (EM)}\label{subsubsec:gs}
In this paper, we extended the implementation of GS proposed by Stolfi et al.~\cite{STOLFI2018250} for encoding the dynamic driving environment of the \avut. \textit{GS} regulates gene expression in a cell to prevent the expression of a certain gene through DNA methylation~\cite{bender2004dna}. As illustrated in Figure~\ref{alg:nbr}(c), the core of \method's \textit{GS} is \textit{GS Probability Generation (EG)}, which generates GS probabilities from the environment. 
GS probabilities ($Pr_{gs}$) is a vector of probabilities
indicating the likelihood of changing each gene/position in solution $x$. As shown in Algorithm~\ref{alg:gs}, the expression of each gene in $x$ is the effect of the nucleosome $n$, the epigenetic probability $Pr_e$, and GS probabilities $Pr_{gs}$. $Pr_e$ is a hyperparameter of \epiga, which controls whether GS takes effect. For example, only if position $k$ in $n$ is collapsed, $x(k)$'s value has a probability of $Pr_{gs}(k)$ to be changed.

Different from the original \epiga, which uses equal GS probabilities (i.e., 0.5) for all the positions, \method is embedded with a machine learning model, namely \textit{Epigenetic Model}, to generate GS probabilities from the environment. We build the model based on a multi-modal fusion transformer~\cite{9863660} and attention mechanism. The model receives observations from the environment and generates GS probabilities accordingly. The details of \textit{EG} will be introduced in Section~\ref{subsec:model}.


\subsection{Attention-based GS Probability Generation}\label{subsec:model}
\begin{figure*}[htbp]
  \centering
  \includegraphics[width=\linewidth]{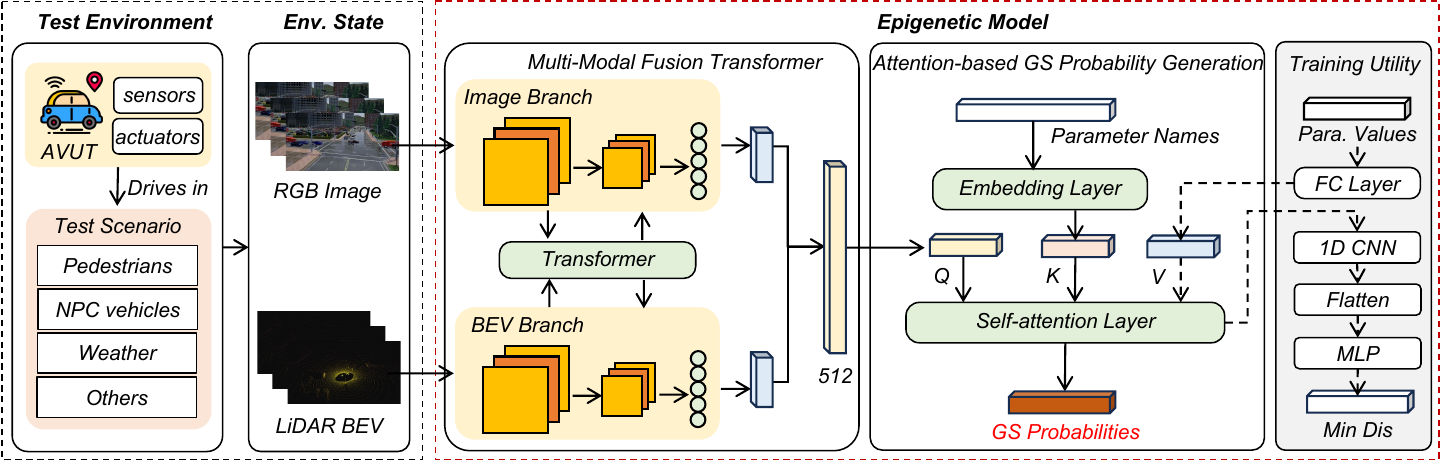}
  \caption{\textbf{Overview of GS Probability Generation}.}
  \label{fig:eg_model}
\end{figure*}

As discussed in Section~\ref{subsubsec:gs}, \method adopts GS as its epigenetic mechanism, where a machine learning model called \textit{Epigenetic Model} is employed to generate GS probabilities. Concretely, as Figure~\ref{fig:eg_model} shows, \textit{Epigenetic Model} receives states/observations (i.e., RGB image and LiDAR bird's-eye view (BEV)) from the test environment and adaptively generates GS probability for each configurable environment parameter. Adaptively doing this, rather than uniformly and statically assigning an equal probability to all parameters, stems from the dynamic and continuously evolving nature of the operating environment of \avut, which consequently exerts distinct effects on the behavior of the \avut~\cite{8667866}. 

Considering that the \avut senses the test environment from multiple data modalities such as images and LiDAR point clouds, we adopt a \textit{Multi-Modal Fusion Transformer}~\cite{9863660} (Section~\ref{subsubsec:mmft}) to fuse image and LiDAR point clouds as encodings of environmental states. The transformer extracts high-level features from environmental states, and then the extracted features are inputted into \textit{Attention-Based GS Probability Generation} (Section~\ref{subsubsec:aeg}) to calculate the GS probabilities. Besides, to train the model, we employ another module named \textit{Training Utility} (Section~\ref{subsub:tu}). 

\subsubsection{Multi-Modal Fusion Transformer (MMFT)}\label{subsubsec:mmft}
To design \textit{MMFT}, we take inspiration from the transformer-based sensor fusion architecture proposed in \textit{Transfuser}~\cite{9863660}. \textit{Transfuser} is a novel model architecture for end-to-end autonomous driving, and it is ranked fourth place in autonomous driving on CARLA leaderboard\footnote{\url{https://leaderboard.carla.org/}}, demonstrating its state-of-the-art sensor fusion ability.
\textit{MMFT} takes RGB and BEV images as its input and outputs a 48-dimensional feature vector representing the state encodings of the environment.
Specifically, we follow the same input representations that use three camera images (front, left, right) and compose them into a single three-channel RGB image with $960 \times 160$ pixels. The BEV image is represented as a three-channel pseudo-image of size $336 \times 150$ pixels. The \textit{Image Branch} and \textit{BEV Branch} are designed as two regular networks (RegNet)~\cite{radosavovic2020designing} with the same network structures employed by \textit{Transfuser}.
As for \textit{Transformer}, different from \textit{Transfuser} that employs four transformer modules, we use one module because we experimented with various settings and found that one module has already achieved comparable prediction results with less training time.

\subsubsection{Attention-Based GS Probability Generation (AGPG)}\label{subsubsec:aeg}
As shown in Figure~\ref{fig:eg_model}, \textit{AGPG} contains two layers: \textit{Embedding Layer} and \textit{Self-attention Layer}.

\textbf{\textit{Embedding Layer (Emb)}} maps discrete input tokens such as words to continuous embedding vectors in a high-dimensional space, where the vectors are a representation of the semantics of tokens, efficiently encoding semantic information that might be relevant to the task at hand~\cite{levy2014dependency}. As Figure~\ref{fig:eg_model} shows, it accepts a 10-dimensional vector (i.e., \textit{Para. Names}) of strings encoding the names of the configurable environment parameters as its input and outputs word embeddings as $\textbf{K}\in\mathbb{R}^{N_k \times D_k}$, where $N_k$ is the number of parameter names and each name is represented as a 1-dimensional feature vector of size $D_k$.

\textbf{\textit{Self-attention Layer (Attn)}} calculates the attention weights between the state encoding and the word embeddings as the GS probabilities.
Concretely, we exploit the self-attention mechanism to map a set of queries ($\textbf{Q}\in \mathbb{R}^{N_q\times D_q}$), keys ($\textbf{K}\in \mathbb{R}^{N_k\times D_k}$), and values ($\textbf{V}\in \mathbb{R}^{N_v\times D_v}$) to outputs, where \textbf{Q} is calculated as the state encoding by \textit{MMFT}, and \textbf{K} are the word embeddings of the parameter names, and \textbf{V} is calculated by the \textit{FC Layer} employed in \textit{Training Utility}:
\begin{equation}
    \textbf{Q}=\textsc{Mmft}(RGB, BEV), \textbf{K} = \textsc{Emb}(\textit{PN}), \textbf{V}=FC(\textit{PV}).
\end{equation}
We then calculate the attention weights \textbf{A} using the scaled dot products between \textbf{Q} and \textbf{K}:
\begin{equation}
    \textbf{A} = sigmoid(\dfrac{Q\cdot K^T}{\sqrt{d_k}}).
\end{equation}
%
As explained in Section~\ref{subsec:attention}, we employ the \textit{sigmoid} as the activation function, and 
the scaling factor $1/\sqrt{d_k}$ is introduced to counteract the effect of having the dot products grow large in magnitude.

Notice that in our problem, $\textbf{Q}\in \mathbb{R}^{1\times 48}$, $\textbf{K}\in \mathbb{R}^{10\times 48}$, and $\textbf{V}\in \mathbb{R}^{1\times 10}$, therefore $\textbf{A}\in \mathbb{R}^{1\times 10}$, which is a vector of 10 values. For a value $\textbf{A}_i$ in position $i$, it can be denoted as the dot product between the state encodings and the word embedding of the $i$ parameter name: $\textbf{Q} \cdot K_i^T$ indicating $ith$ parameter's contribution to the output.
The attention weights are continuously optimized as the training proceeds, and after the model has been trained, we use the attention weights as 
probabilities of changing/silencing each gene in a solution, i.e., GS probabilities.

In addition, the output of the attention layer is denoted as the concatenation of \textbf{Q}, the dot products between \textbf{A} and \textbf{V} ($\textbf{A} \cdot \textbf{V}$), and \textbf{V}:
\begin{equation}
    \textbf{O}_\textbf{attn} = concatenate(A \cdot V, Q, V) = A \cdot V + Q + V,
\end{equation}
where $\textbf{A} \cdot \textbf{V}$ is computed as a weighted sum of the values of \textbf{V}, whereas \textbf{Q} and \textbf{V} are directly connected to the output, referred to as shortcut connections. Shortcut connection is a commonly applied technique in Residual Networks~\cite{he2016deep}, which has been shown effective in preventing vanishing/exploding gradients. For training the network, $\textbf{O}_\textbf{attn}$ is further passed into \textit{Training Utility}.

\subsubsection{Training Utility}\label{subsub:tu}
This module trains \textit{Epigenetic Model}. Once the model is well-trained, it is used for generating GS probabilities. The training utility's architecture is shown in Figure~\ref{fig:eg_model}. Specifically, it takes a 10-dimensional vector representing the values of the configurable environmental parameters (\textit{Para. Values}) as its input and passes it to the \textit{FC Layer} with 10 neurons. The output of the \textit{FC Layer} is passed to the \textit{Self-attention Layer} in \textit{AGPG} (Section~\ref{subsubsec:aeg}) to get $\textbf{O}_\textbf{attn}$. $\textbf{O}_\textbf{attn}$ is a 106-dimensional vector, and we pass it to a 1-dimensional convolutional (Conv1D) network (i.e., \textit{1D CNN}). The \textit{1D CNN} has four Conv1D layers with the kernel sizes being 16, 32, 64, and 32, respectively. The output of \textit{1D CNN} is then passed to the \textit{Flatten Layer} and will be flattened into a 3136-dimensional vector. Finally, the output is passed to the multi-layer perception (\textit{MLP}), which predicts the fitness value (i.e., the minimum distance as discussed in~\ref{subsec:optimization_pro}). The \textit{MLP} contains four fully connected layers with the neuron numbers 512, 256, 128, and 1, respectively.
We calculate smooth \textit{L1} loss between the predicted fitness value $f$ and ground truth fitness value:

\begin{equation}
    \mathcal{L} = \begin{cases}
    \dfrac{1}{n}\sum_{i=1}^{n}\dfrac{1}{2}({f_{gt}}_i-f_i)^2, |{f_{gt}}_i-f_i| < 1\\
    \\
    \dfrac{1}{n}\sum_{i=1}^{n}(|{f_{gt}}_i-f_i|-0.5), \text{otherwise}
    \end{cases}
\end{equation}

\section{Experiment Design}\label{sec:experiment_design}

\subsection{Research Questions}\label{subsec:RQs}
We evaluate \method by answering the following three research questions (RQs). 
\begin{compactitem}
    \item RQ1: How effective is \method compared to the two baseline methods?
    \item RQ2: How efficiently does \method perform compared to the baseline methods?
    \item RQ3: How does gene silencing affect the gene expression (i.e., selection of parameters) in \method and \methodequal?
\end{compactitem}

\subsection{Subject System and Simulator}\label{subsec:system_simulator}
To evaluate \method, we employed Interfuser~\cite{shao2023safety} as the system under test, which is an end-to-end \ads ranked the first place on the CARLA leaderboard. Interfuser has been evaluated in various driving situations from the CARLA public leaderboard, the Town05 benchmark~\cite{prakash2021multi}, and the CARLA 42 Routes benchmark~\cite{chitta2021neat}, demonstrating its outstanding performance in comprehensive scene understanding and adversarial event detection. 

For the simulator, we adopted CARLA~\cite{dosovitskiy2017carla}, an open-source and widely-applied autonomous driving simulator, to simulate the \avut and its driving environment. CARLA provides an extensive list of digital assets, including vehicles, pedestrians, sensors, and high-definition maps, for supporting the development, training, and validation of AVs.
In our experiments, we used CARLA-0.9.10.1 and its default Interfuser settings. The \avut controlled by Interfuser is the Tesla Model 3, which has been used in various autonomous driving research contexts~\cite{endsley2017autonomous}.

\subsection{Baselines}\label{subsec:baselines}
As Arcuri and Briand suggested~\cite{6032439}, random strategies (\textit{RS}) are usually used for sanity checks; therefore, to check if our problem is complex, we performed a pilot study to compare \method with \textit{RS}. Results show that \method is significantly better than \textit{RS} in collision scenario generation. Hence, our formal experiment excluded \textit{RS} as a baseline. The detailed results of the pilot study are provided in our online repository (see Section~\ref{subsec:data}).

To answer RQs, we employed two baselines, including a classical GA (i.e., \ga)~\cite{mirjalili2019genetic} without any epigenetic mechanism implemented, and a modified \method, i.e., \methodequal, which sets equal GS probabilities for each gene, i.e., environmental parameters.
\textit{\textbf{\ga}} is a widely applied, single-objective metaheuristic algorithm that uses biologically inspired genetic operators such as mutation, crossover, and selection to solve search problems. For our experiments, we implemented \ga using jMetalPy~\cite{BENITEZHIDALGO2019100598}, a well-known framework for single/multi-objective optimization. \textbf{\textit{\methodequal}} uses equal GS probabilities for each gene, i.e., 0.5, the default setting employed by \epiga~\cite{STOLFI2018250}. This means that each gene has the same chance to be silenced or expressed, which further implies that each gene is treated equally,
and for each gene, the probability of being activated and silenced is equal to 0.5.


\subsection{Parameter Settings}\label{subsec:parameter_settings}

Since the parameters of \method, \methodequal, and \ga greatly impact the algorithm's performance, finding appropriate parameter settings for genetic and epigenetic algorithms is crucial to obtaining optimal results and fair comparisons. Therefore, to determine the parameters for \ga and \epiga employed by \method and \methodequal, we experimented with different combinations of the key parameters. As a result, we set the population size as 20 and the number of cells in each individual of \method and \methodequal as 1. Besides, the termination criterion is the maximum number of evaluations, i.e., 1000. In addition, we set the nucleosome probability $Pr_n$ as 0.2, the nucleosome radius $R$ as 1, and the epigenetic probability $Pr_e$ as 0.01 (Section~\ref{subsubsec:ng_nbr}).

As for hyperparameter settings for training the epigenetic model of \method, we set the batch size and number of epochs to 64 and 10000 based on the results of our pilot study. For the other hyperparameters, we followed the same settings as the model from Transfuser. To train the epigenetic model, we first built a dataset by running a random strategy. As a result, we obtained a dataset containing about 25k samples, each labeled with a fitness value, i.e., minimum distance. Since the epigenetic model aims to calculate attention weights to represent the GS probability for each gene that can potentially contribute to collisions or safety violations of the \avut, samples with higher chances (i.e., smaller minimum distances) of causing safety violations are more important. Therefore, we filter out records in the dataset with a minimum distance greater than 5 meters. In the end, we obtained a dataset of around 5000 samples. We follow the 80\%/20\% training/test split ratio to split the dataset into a training and a test dataset, as suggested by R{\'a}cz et al.~\cite{racz2021effect}. The model was trained on the training dataset and converged with a loss value 0.019. We evaluated the trained model on the test dataset and obtained a mean square error (i.e., MSE) of 0.027 and a mean absolute error (i.e., MAE) of 0.16, indicating the prediction error for the minimum distance is about 0.16m, which is acceptable since the minimum distance ranges from 0 to 5m.
The pilot study results, the epigenetic model parameter settings, and training details are available in our online repository (see Section~\ref{subsec:data}).

\subsection{Test Environment Initialization and Execution}
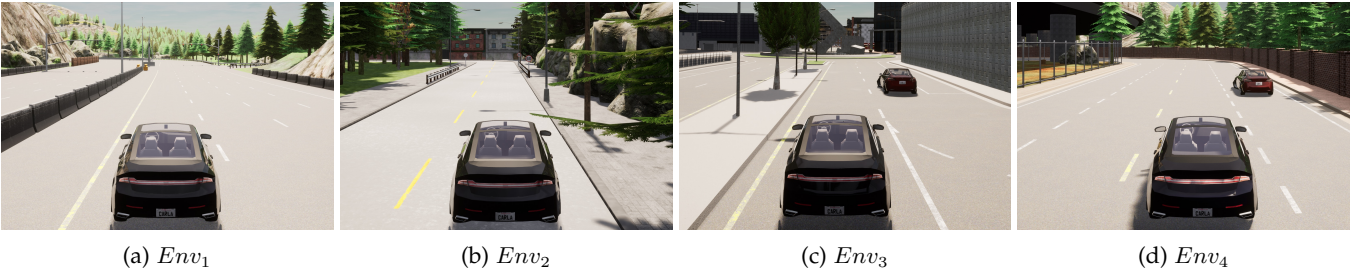
\begin{figure*}[htbp] 
	\centering
        \resizebox{\textwidth}{!}{
        \input{figures/scenarios}
        }
	\caption{\textbf{Initial Driving Environments of the \avut.} \textbf{Sub-figure~\ref{fig:s1}} depicts $Env_1$ where the \avut is driving along a one-way road with four lanes of the Town04 map. The \avut is expected to change lanes twice from the leftmost lane to the rightmost and then maintain the lane until arriving at its destination;
 \textbf{sub-figure~\ref{fig:s2}} depicts $Env_2$ where the \avut is driving along a dual-way road with two opposite lanes of the Town2 map. The \avut is expected to maintain its current lane until arriving at its destination;
 \textbf{sub-figure~\ref{fig:s3}} depicts $Env_3$ where the \avut is driving along a dual-way road with two opposite lanes and then entering into a one-way road with two lanes of the Town03 map. Meanwhile, a red sedan is trying to switch from its current lane to the left lane; 
 \textbf{sub-figure~\ref{fig:s4}} depicts $Env_4$ where the \avut is driving along a dual-way four-lane road with two lanes in each direction of the Twon05 map. Meanwhile, a red sedan is trying to switch from its current lane to the left lane. 
    }
    \label{fig:env}
\end{figure*}

\method identifies critical scenarios by configuring an initial test environment, i.e., introducing a \npc and a pedestrian, and manipulating the weather as discussed in Section~\ref{subsec:optimization_pro}. In our design, an initial test environment specifies the map, the \avut's driving route, and existing traffic users such as vehicles.
The driving route specifies that the \avut should drive from a starting point to a destination without collisions. Notice that in the initial environment, the \avut can navigate safely to the destination if no additional environmental scenario element (e.g., the \npc and weather) is introduced, and \method's goal is to introduce new environmental scenario elements that can cause collisions or other safety violations. 
In our experiment, we select four initial environments, i.e., $Env_1$, $Env_2$, $Env_3$, and $Env_4$, as described in Figure~\ref{fig:env}.

To account for the randomness caused by the \avut and the simulator we used, 
we executed \method, \ga, and \methodequal 10 runs on each initial environment. 
After each run, we obtained a solution (i.e., test input), and we further executed it 30 times to deal with the \avut and simulator randomness. Finally, we obtained results of 3600 executions (3 methods $\times$ 4 initial environments $\times$ 10 runs $\times$ 30 executions).

All experiments were executed on one server node with an Intel Xeon Platinum 8186, 8$\times$NVIDIA A100 GPU.

\subsection{Evaluation Metrics and Statistical Tests}\label{subsec:metrics}
To evaluate the performance of \method and the baselines, we adopt five metrics commonly used in AV development and testing~\cite{9863660,8453180}:
\begin{asparaenum}
    \item \textit{\textbf{Minimum Distance (\md)}} measures the minimum distance between the \avut and its surrounding objects. A lower \md value indicates a higher chance of safety violations. For the $kth$ execution of solution $s$, \md is calculated as: $MD_{s}^{k} = \mathcal{F}_s(\avut, obj_{s})$, where $F_s$ is the fitness function (Section~\ref{subsec:optimization_pro}).
    \item \textit{\textbf{Collision (\co)}} indicates whether the \avut collided with any object. For the $kth$ execution of solution $s$, $CO_s^k \in \{0, 1\}$, where 0 and 1 denote no collision occurred and a collision happened, respectively;
    \item \textit{\textbf{Route Completion (\rc)}} defines the percentage of route distance completed by the \avut. $RC_s^k$ denotes the route completion of the $kth$ execution of solution $s$. A lower $RC_s^k$ indicates $s$ caused the \avut to complete less route distance, e.g., caused by collisions or traffic jams. 
    \item \textit{\textbf{Infraction Score (\is)}} calculates geometric series of infraction penalty coefficients, with lower \textit{IS} values indicating more serious infractions occurred. \textit{IS} for the $kth$ execution of solution $s$ is calculated as:
    \begin{equation}
        IS_s^k = \prod_j^{ped, npc, sta}(p^j)^{(\#infractions^j)}
    \end{equation}
    where $p_j$ is the penalty coefficient of infraction $j$, and $\#infractions^j$ denotes the number of $infractions^j$ occurred. We consider three types of infraction, i.e., collision with a pedestrian ($ped$), collision with a \npc ($npc$), and collision with a static object ($sta$). According to the CARLA leaderboard, the penalty coefficient for each infraction is 0.50, 0.60, and 0.65, respectively.
    \item \textbf{\textit{Driving Score (\ds)}}, as a comprehensive metric, calculates the weighted route completion (\textit{RC}) with infraction score (\textit{IS}). A lower \textit{DS} value indicates poorer \avut's overall driving performance. \textit{DS} for the $kth$ execution of solution $s$ is calculated as the product of \textit{RC} and \textit{IS}: $DS_s^k = RC_s^k \cdot IS_s^k$.
\end{asparaenum}

\textbf{\textit{Statistical Test}}.
Based on the guidelines~\cite{6032439}, we first perform the Mann-Whitney U test with a significance level of 0.05 to study the statistical significance of two methods and then use the Vargha and Delaney effect size to calculate $\hat{A}_{12}$. The effect size $\hat{A}_{12}$ indicates the chance of method \textit{A} yielding higher values of a metric $\chi$ than method \textit{B}. If $\hat{A}_{12}$ is greater than 0.5, then \textit{A} has a higher chance to obtain higher values of $\chi$ than \textit{B}, and vice versa. 

To study the correlation between \textit{MD} and other metrics (i.e., \textit{CO}, \textit{RC}, \textit{IS}, and \textit{DS}), we perform the Spearman's rank correlation ($\rho$) test, a non-parametric test that measures the monotonic relationship between two ranked variables. $\rho \in (0, 1]$ indicates a positive correlation and $\rho \in [-1, 0)$ shows a negative correlation. A value of 1.0 (-1.0) indicates a perfect positive (negative) correlation, and 0 means no correlation. The significance of a correlation is indicated with a p-value less than 0.05. Based on the guidelines by Mukaka~\cite{mukaka2012guide}, we further divide $\rho$ into five levels to interpret the magnitude of a correlation: \textit{negligible} ($\rho$ $\in$ (-0.300, 0.300)), \textit{low} ($\rho$ $\in$ [0.300, 0.500) or (-0.500, -0.300]), \textit{moderate} ($\rho$ $\in$ [0.50, 0.700) or (-0.700, 0.500]), \textit{high} ($\rho$ $\in$ [0.700, 0.900) or (-0.900, -0.700]), and \textit{very high} ($\rho$ $\in$ [0.900, 1.000] or [-1.000, -0.900]).

\section{Experiment Results and Analyses}\label{sec:results}

\subsection{Results for RQ1 - Effectiveness}\label{subsec:result_rq1}
To answer RQ1, we compared the effectiveness of \method with \methodequal and \ga regarding all metrics under the four initial driving environments (i.e., $Env_1$, $Env_2$, $Env_3$, and $Env_4$) and across them (i.e., $Env_{1-4}$). 

\textbf{\textit{Statistical Differences.}} Results of the Mann and Whitney U test and Vargha and Delaney effect size are reported in Table~\ref{tab:table_1}.
\begin{table*}[htbp]
    \centering
    \small
    \caption{\textbf{Results of pair-wise comparisons of \method with the baseline methods (i.e., \ga and \methodequal) using the Vargha and Delaney statistics and the Mann–Whitney U test -- RQ1}. "$\downarrow$ / $\uparrow$" denotes that a smaller/larger metric value indicates a better performance of a method. A bold $\hat{A}_{12}$ with a \textit{p-value} $<$ 0.05 implies that \method is significantly better than \ga/\methodequal. A $\hat{A}_{12}$ with symbol "×" indicates that \method significantly underperformed \ga/\methodequal.}
    \begin{threeparttable}
    \resizebox{\textwidth}{!}{

\input{tables/table1}
    }
    \end{threeparttable}
    \label{tab:table_1}
\end{table*}
Regarding \md, \method significantly outperformed \ga for all four initial environments, i.e., $\hat{A}_{12} < 0.5$ and $\textit{p-value} < 0.05$. 
\method also significantly outperformed \methodequal in $Env_1$ and $Env_2$ and achieved comparable performance with \methodequal for $Env_3$ and $Env_4$. 
Similarly, for \co, \method significantly outperformed \ga for all initial environments regarding leading the \avut to collide
and significantly outperformed \methodequal in $Env_1$ and $Env_2$ and achieved comparable performance with \methodequal in $Env_3$ and $Env_4$.

Regarding \rc, we can observe that \method performed significantly better (i.e., lower route completion) than \ga in $Env_4$, while it significantly underperformed \ga in $Env_1$, $Env_2$, and $Env_3$. \method significantly outperformed \methodequal in $Env_1$, $Env_3$, and $Env_4$ and underperformed in $Env_2$. These observations show that, in most cases, \method can cause the \avut to collide earlier and, therefore, complete less route distance than \methodequal and \ga.
After replaying generated scenarios, we observed cases that \method led the \avut to collide, but it continued to drive forward. Hence, \rc further increases. However, \ga generated scenarios can cause the \avut to get stuck in traffic jams or blocked by its front obstacles, achieving better performance in \rc. We also observed a similar situation in $Env_2$ for \methodequal. Therefore, in \method, low \rc values are mainly caused by collisions, not unrealistic traffic jams, etc.

Regarding \is, \method outperformed \ga in all initial environments, and for $Env_1$, $Env_2$, and $Env_3$ the differences are significant, i.e., \textit{p-value}$<$0.05. \method performed significantly better than \methodequal in $Env_1$ and $Env_2$, while there is no significant difference in $Env_3$ can be observed. However, in $Env_4$, \method significantly underperformed \methodequal. Recall that \is calculates the geometric series of penalty coefficients by considering three types of collisions: with dynamic objects (pedestrian and \npc) and with static objects. Each type is associated with a penalty coefficient indicating the collision's severity, i.e., the smaller the coefficient, the more severe the collision. A higher penalty coefficient leads to a higher \is, which is computed by multiplying a penalty coefficient for every collision. Therefore, the result in $Env_4$ indicates that \methodequal caused more collisions with lower penalty coefficients than \method.
To know why, we further analyzed scenarios that led to collisions and found that \methodequal caused more collisions with pedestrians whose penalty coefficient is 0.5, while most collisions caused by \method are with NPC vehicles whose penalty coefficient is 0.6. In CARLA leaderboard, collisions with pedestrians are considered more severe than those with NPC vehicles, so collisions with pedestrians have a lower penalty coefficient than those with NPC vehicles. 
However, there is evidence that collisions with trucks can cause more severe injuries to passengers than collisions with pedestrians~\cite{DESAPRIYA200442}.

Regarding \ds, \method achieved comparable performance with \ga in $Env_1$, $Env_2$, and $Env_3$ and significantly outperformed \ga in $Env_4$. \method significantly outperformed \methodequal in $Env_1$ and $Env_4$ and has no significant difference observed in $Env_2$ and $Env_3$. As a performance metric calculated by weighting \rc and \is, the results for \ds show that the \avut performed similarly or significantly worse in scenarios generated by \method than in those generated by \ga and \methodequal.

Besides, when looking at the results by combining all four initial environments, i.e., row $Env_{1-4}$ in Table~\ref{tab:table_1}, we can observe that, overall, \method outperformed \ga and \methodequal in terms of all metrics, and the results of \method are all significantly better except for the comparison with \methodequal regarding \rc.

\textit{\textbf{Correlation Analysis among the Five Metrics.}} Considering that \md is the objective that \method directly optimizes, we performed the Spearman's rank correlation ($\rho$) test to study the correlation between \md and the other metrics and report the results in Table~\ref{tab:RQ1_cor_test}. 
As the table shows, there is a high/very high and significant negative correlation between \md and \co for all three methods in all four environments, i.e., all $\rho$ values are near -1.0, and a high/very high and significant positive correlation between \md and \is for almost all cases (except for \ga in $Env_4$). 
This is reasonable as the lower the distance between the \avut and other objects, the higher the chance of more collisions and lower infraction scores.
We can also observe negligible correlations between \md and \rc/\ds.

\begin{figure*}[htbp]
    \centering
    \includegraphics[width=\textwidth]{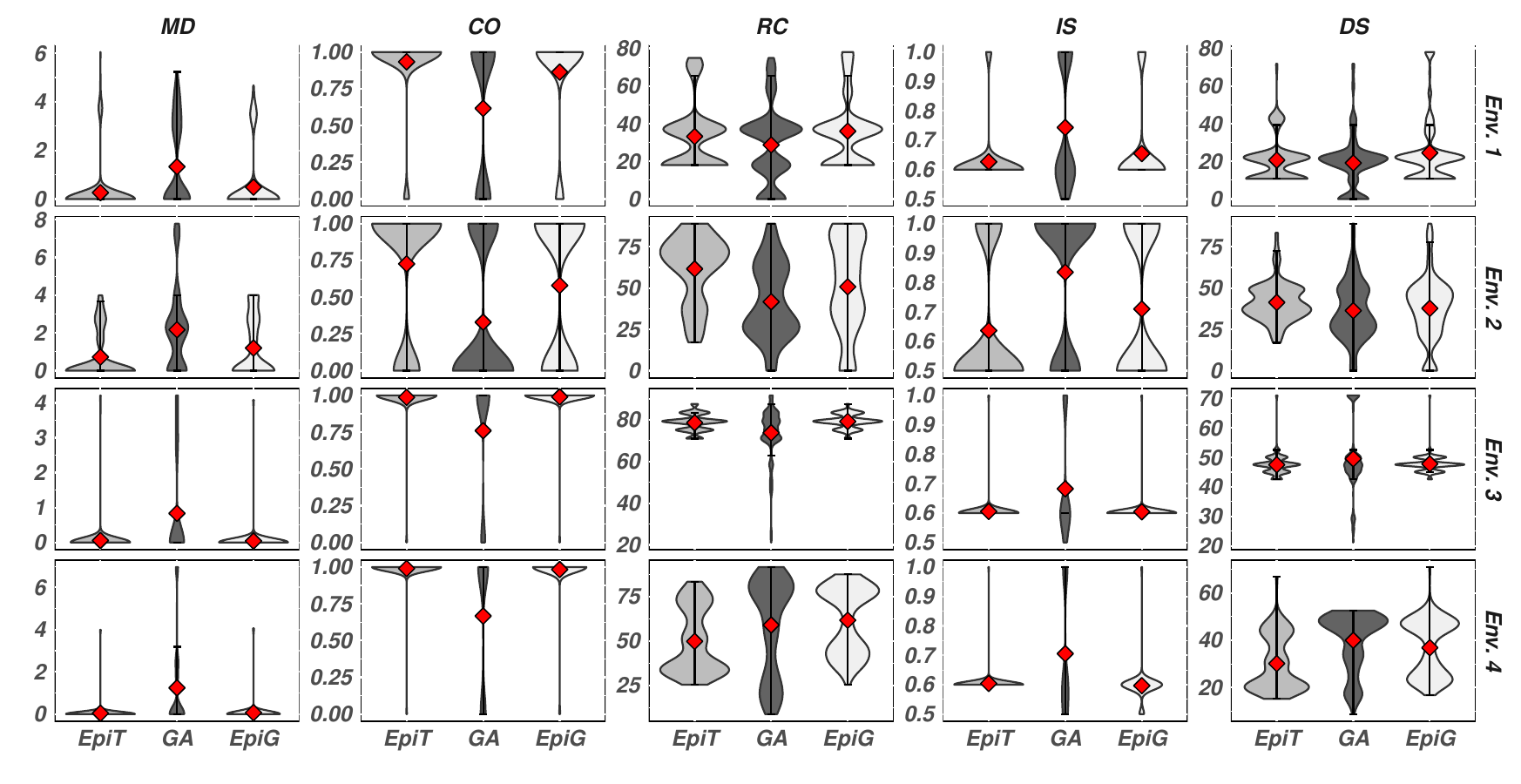}
    \caption{\textbf{Descriptive statistics of all the metrics achieved by \method, \ga, and \methodequal -- RQ1.} \textit{EpiT} is \method and \textit{EpiG} is \methodequal. $\begingroup \color{red} \mdblkdiamond \endgroup$ denotes the mean value of each sample.
    }
    \label{fig:RQ1_box_plot}
\end{figure*}

\textbf{\textit{Distributions.}} We also present the descriptive statistics of results achieved by each method in Figure~\ref{fig:RQ1_box_plot}. Regarding \md, across all initial environments, \method and \methodequal achieved similar distributions of lower variability than those from \ga. Similar patterns on the variability of the \co and \is distributions achieved by \method and \methodequal can be observed. This indicates that GS employed in both \method and \methodequal is reliable in directly reducing \md (as it is the optimization objective) and indirectly increasing \co and reducing \is. 
For \rc and \ds, the distributions obtained by \method and \methodequal also have lower variability than those from \ga. When looking across the different metrics, the results on \rc and \ds are less reliable than those on \md, \co, and \is. This observation further confirms the results of the correlation analyses.


\begin{table}[htbp]
    \centering
    \caption{\textbf{Results of correlations between \textit{MD} and \textit{CO}, \textit{RC}, \textit{IS}, and \textit{DS} using the Spearman's rank correlation ($\rho$) test}. A $\rho$ value decorated by a $\begingroup\color{blue}\blacktriangle\endgroup$ indicates that \textit{MD} has a high/very high and significant positive/negative correlation with another metric, i.e., $\rho$ $\in$ [0.700, 1.000] or [-1.000, -0.700] and $\textit{p-value} < 0.05$. A $\rho$ value decorated with a $\bullet$ indicates insignificant correlation between \textit{MD} and another metric.}
    \resizebox{\columnwidth}{!}{

\input{tables/cor_test}
    }
    \label{tab:RQ1_cor_test}
\end{table}

\begin{center}
    \fcolorbox{black}{gray!10}{
    \parbox{0.96\columnwidth}{
    \textbf{Conclusion for RQ1}: 
    Compared to \methodequal and \ga, \method achieved the \textit{overall} best performance regarding the selected metrics, and the differences are mostly significant.
    This indicates that \method is more effective in generating critical scenarios that lead to shorter distances to other objects, more collision occurrences, and poorer driving performance of the \avut, e.g., lower driving score. 
    }}
\end{center}

\subsection{Results for RQ2 - Efficiency}\label{subsec:result_rq2}
\begin{figure*}[htbp]
    \centering
    \includegraphics[width=\textwidth]{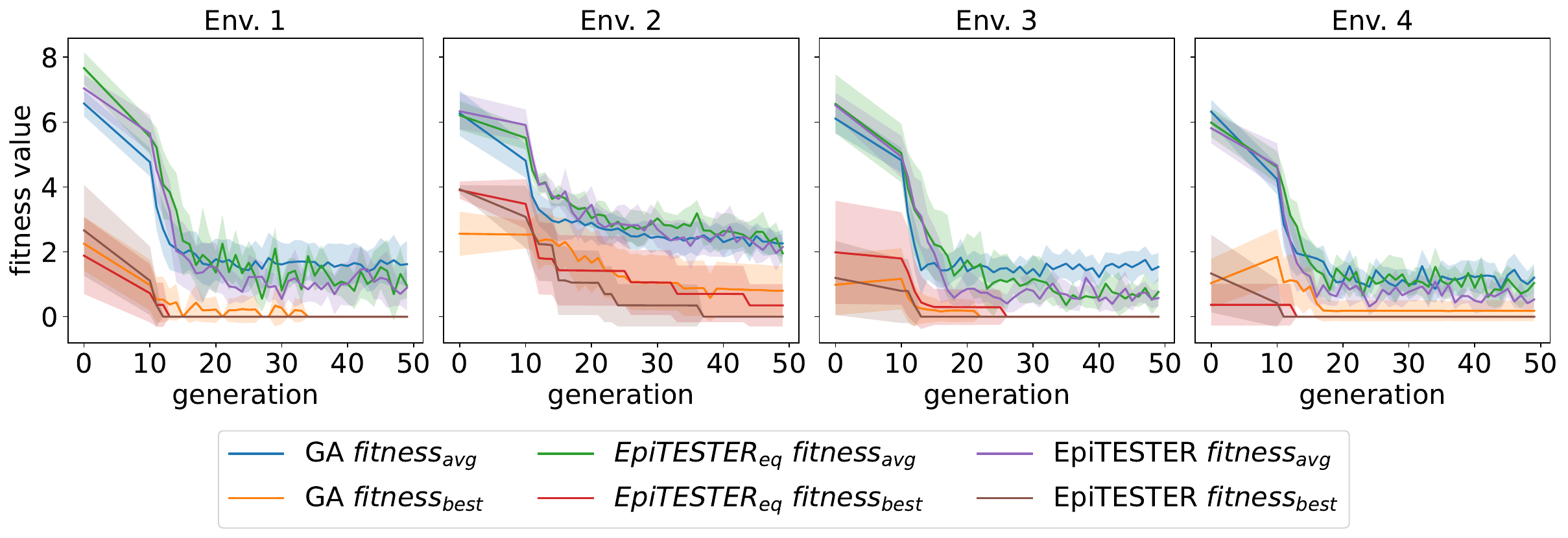}
    \caption{\textbf{Average and best fitness values (i.e., $fitness_{avg}$ and $fitness_{best}$) achieved by \method, \ga, and \methodequal with 95\% confidence intervals in each generation -- RQ2.} 
    }
    \label{fig:conv}
\end{figure*}

Recall that we set the population size as 20 and the number of generations as 50; therefore, the number of evaluations is 1000 for each method. We ran each method 10 times to account for the randomness. Therefore, for each run $r$ ($r$=1...10) of each method, we obtained 50 generations, and each has 20 solutions. We then calculate the average and the best fitness values (i.e., \md) for each generation of 20 solutions as $fitness_{avg}$ and $fitness_{best}$. 
Figure~\ref{fig:conv} presents how $fitness_{avg}$ and $fitness_{best}$ of the 10 runs vary in each generation of \method, \ga, and \methodequal in each initial environment. Concretely, we report the mean values of $fitness_{avg}$ and $fitness_{best}$ of each generation with 95\% confidence intervals.

As shown in Figure~\ref{fig:conv}, \ga converged to higher $fitness_{avg}$ values than \method and \methodequal in all initial environments except for $Env_2$ where all three methods converged to a comparable $fitness_{avg}$. Besides, \method and \methodequal converged to a similar $fitness_{avg}$ in all initial environments, and in $Env_4$, \method consistently performed better than \methodequal throughout the 50 generations. When looking at the evolution along the generations, compared to \methodequal, \method achieved faster convergence speeds in $Env_1$, $Env_3$, and $Env_4$, and a comparable convergence speed with \methodequal in $Env_2$. This indicates that \method is more efficient at minimizing the distance between the \avut and other objects than \methodequal. \ga quickly converged in the first 10 to 20 generations and then gradually slowed down. When comparing the variability of $fitness_{avg}$ achieved by the methods, we do not notice many differences, implying that they have comparable reliability.

As for $fitness_{best}$ in each generation of the three methods, we can see from Figure~\ref{fig:conv} that for all 10 runs, \method can always find 0 as $fitness_{best}$ in all initial environments when converged, while \ga and \methodequal only find 0 as $fitness_{best}$ in $Env_1$ and $Env_3$ for all 10 runs. Besides, when looking at the number of generations needed to detect a collision (i.e., $fitness_{best}$=0), it is evident that \method used fewer generations than \ga and \methodequal, in all four environments, meaning that \method is more efficient at detecting collisions. One can also notice that the variances of $fitness_{best}$ achieved by the three methods are all large initially and then smaller over generations. As for \method, we can always observe no variability in terms of $fitness_{best}$ when it converged for all initial environments, while we can observe two such cases for \ga (i.e., $Env_1$ and $Env_3$), and three such cases for \methodequal (i.e., $Env_1$, $Env_3$, and $Env_4$).



\begin{center}
    \fcolorbox{black}{gray!10}{
    \parbox{0.96\columnwidth}{
    \textbf{Conclusion for RQ2}:
    Compared to \methodequal and \ga, \method achieved a faster convergence speed and used fewer generations to detect collisions.
    This shows that compared to the baseline methods, the \method is not only more efficient at converging to shorter distances between \avut and other objects but also more efficient at collision detection.
    }}
\end{center}

\subsection{Results for RQ3 - Gene Silencing}
\begin{figure*}[htbp]
    \centering
    \includegraphics[width=\textwidth]{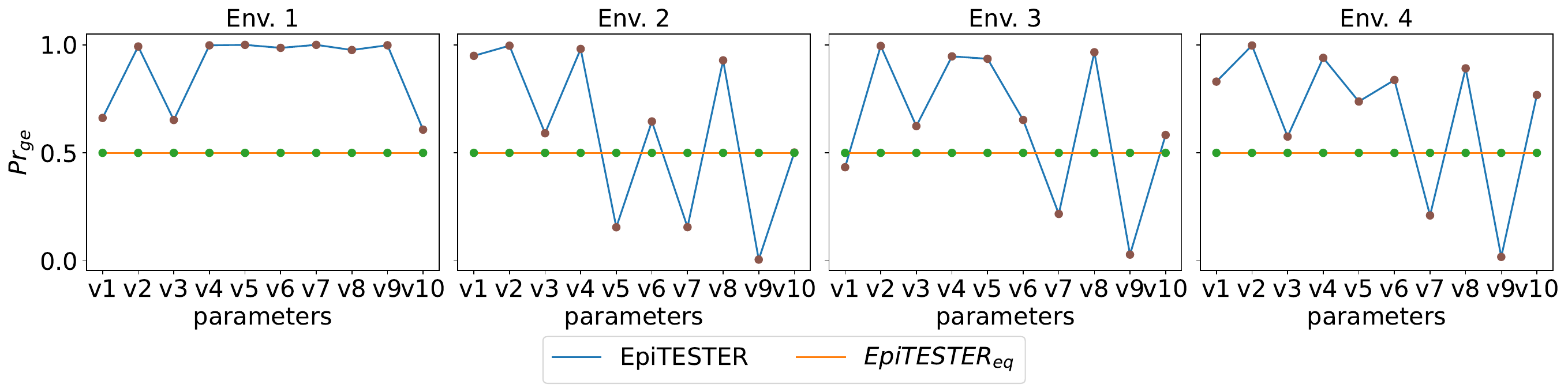}
    \caption{
    \textbf{Gene expression probabilities (i.e., $Pr_{ge}$) predicted by the epigenetic model in \method (blue) and the default $Pr_{ge}$ applied in \methodequal (orange) -- RQ3.}
    v1: \textit{$dis^{lo}_{npc}$}, v2: \textit{$dis^{la}_{npc}$}, v3: $behaviour_{npc}$, v4: \textit{$dis^{lo}_{ped}$}, v5: \textit{$dis^{la}_{ped}$}, v6: \textit{$o^x_{ped}$}, v7: \textit{$o^y_{ped}$}, v8: $v_{ped}$, v9: \textit{$angle_{sun}$}, v10: $density_{fog}$.
    }
    \label{fig:attention}
\end{figure*}

\begin{figure*}[htbp]
    \centering
    \includegraphics[width=\textwidth]{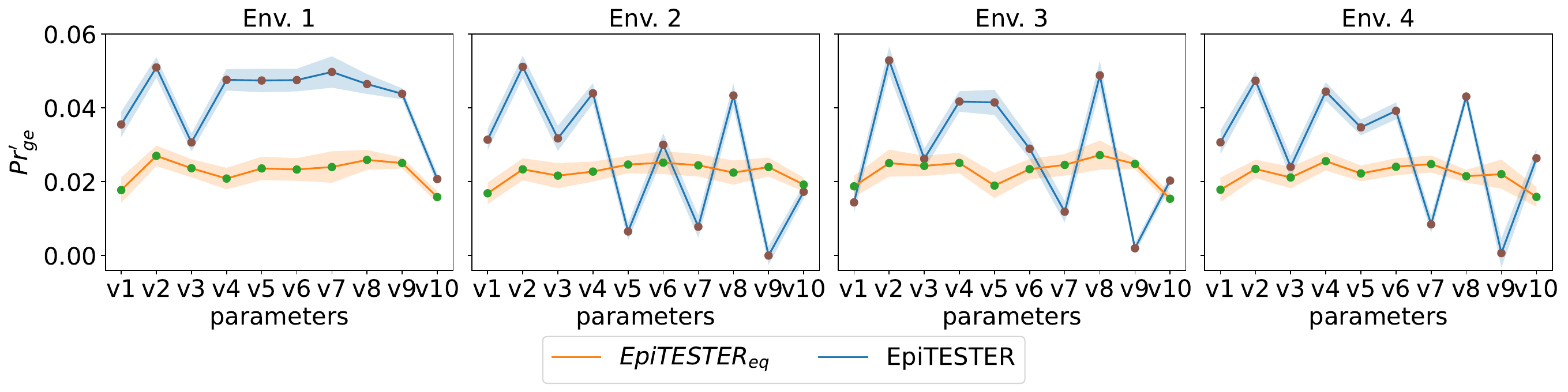}
    \caption{\textbf{Gene expression probability distributions  (i.e., $Pr_{ge}'$) of \method and \methodequal with 95\% confidence intervals -- RQ3.} 
    v1: \textit{$dis^{lo}_{npc}$}, v2: \textit{$dis^{la}_{npc}$}, v3: $behaviour_{npc}$, v4: \textit{$dis^{lo}_{ped}$}, v5: \textit{$dis^{la}_{ped}$}, v6: \textit{$o^x_{ped}$}, v7: \textit{$o^y_{ped}$}, v8: $v_{ped}$, v9: \textit{$angle_{sun}$}, v10: $density_{fog}$.
    }
    \label{fig:sliencing}
\end{figure*}

To answer RQ3, for each gene (i.e., each of the 10 configurable environmental parameters), we first report its gene expression probability ($Pr_{ge}$) used in \method and \methodequal in Figure~\ref{fig:attention}. The probability $Pr_{ge}$ is equal to one minus the gene silencing probability (i.e., $Pr_{gs}$).
Then we statisticize the actual gene expression probability distributions ($Pr_{ge}'$) achieved by executing \method and \methodequal in each initial environment, and Figure~\ref{fig:sliencing} presents the mean with 95\% confidence intervals of $Pr_{ge}'$ in each environment.


As shown in Figure~\ref{fig:attention}, the gene expression probabilities for each parameter applied in \methodequal are equal to 0.5 in all four environments, as it is the default setting of \methodequal. As for \method, recall that it employs an attention mechanism (Section~\ref{subsubsec:aeg}) to identify a suitable parameter that contributes to safety violations of the \avut by regulating its expression via its gene expression probability, which is generated adaptively along with the environment state changes. 
Hence, we can observe that the gene expression probabilities differ from parameter to parameter. However, specific patterns can be observed across the environments; the parameters (v1-v8) related to the dynamic objects (\npc and pedestrian) tend to have higher probabilities of being expressed than the weather parameters (v9 and v10). This suggests that configuring dynamic object parameters is more likely to lead to safety violations. Furthermore, we can observe that $dis^{la}_{npc}$ always receives the highest expression probability, which is close to 1, for all four environments, meaning that it has the highest contribution to safety violations. This is because $dis^{la}_{npc}$ is related to lateral collisions due to lane-changing behaviors. 

Recall from Algorithm~\ref{alg:gs} that the expression of each parameter is controlled by the nucleosome $n$, the epigenetic probability $Pr_e$, and the GS probabilities $Pr_{gs}$ (i.e., 1-$Pr_{ge}$), where $n$ is a binary mask calculated in the \textit{NG} operator (Section~\ref{subsubsec:ng_nbr}), and $Pr_e$ is a hyperparameter which is set to 0.01. The GS mechanism functions only when a position in $n$ is collapsed, and a randomly generated number is smaller than $Pr_e$; thus, as shown in Figure~\ref{fig:sliencing}, $Pr_{ge}'$ achieved by each method is smaller than $Pr_{ge}$.
However, as suggested in the figure, $Pr_{ge}'$ follows the same pattern as $Pr_{ge}$ in Figure~\ref{fig:attention} over 10 runs. For example, \textit{$dis^{la}_{npc}$} always has the highest chance of being expressed. 
Regarding the variability of $Pr_{ge}'$, \method achieved a lower variability than \methodequal, suggesting that, compared to \methodequal, \method is more confident about which parameter should be expressed or silenced.


\begin{center}
    \fcolorbox{black}{gray!10}{
    \parbox{0.96\columnwidth}{
    \textbf{Conclusion for RQ3}: 
     GS mechanism can affect the expression of environmental parameters, and the attention mechanism can effectively identify the contribution of each parameter to collisions or safety violations as the probability of expressing each parameter. Specifically, parameters related to dynamic objects have higher chances of being expressed than weather parameters.
    }}
\end{center}

\subsection{Threats to Validity}\label{subsec:threat}

\textbf{\textit{Conclusion Validity}} concerns the validity and reliability of conclusions drawn in the empirical study. Since \method is a search-based AV testing method, to account for the inherent randomness of the search algorithms, we repeated each experiment 10 times -- commonly used in other AV testing research~\cite{10.1109/ICSE48619.2023.00155}. We further executed each obtained solution 30 times to account for the inherent randomness in AV. Following the guidelines from~\cite{6032439}, we performed statistical tests to draw solid conclusions.

\textbf{\textit{Construct Validity}} is related to the metrics used to compare \method with the baseline methods. To ensure the fairness of the comparisons, we chose five evaluation metrics commonly used in the context of AV testing and evaluation~\cite{9863660,8453180} and applied them consistently to all methods.

\textbf{\textit{Internal Validity}} concerns our \method's hyperparameter settings. To mitigate the potential threats to the internal validity,
we determined the hyperparameter settings by experimenting with different parameter configurations and chose the best one as our hyperparameter setting. We acknowledge that \method's performance might be further improved with more optimized hyperparameter settings; however, this would require dedicated and large-scale empirical studies. 

\textbf{\textit{External Validity}} is about the generalization of the empirical study. We employed one subject system and one simulator to build the test environment, which can potentially threaten the external validity. To mitigate the issue, we chose the state-of-the-art \ads (i.e., Interfuser) that ranked first on the CARLA leaderboard. We acknowledge that conducting more case studies could strengthen our conclusions, and in the future, we plan to explore more simulators and \ads to strengthen our conclusions.

\subsection{Data Avalibility}\label{subsec:data}
To promote open science, we provide the replication package in an online repository:~\url{https://github.com/Simula-COMPLEX/EpiTESTER}. 
Once the paper is accepted, we will release it in a permanent repository, such as Zenodo.

\section{Discussions}\label{sec:discussions}

\subsection{Benefiting from attention mechanism}
Many environmental parameters can be employed to characterize test scenarios, but not all contribute equally to safety violations. 
Hence, the attention mechanism in \method allows the epigenetic model to dynamically allocate more attention (i.e., lower GS probabilities) to specific parameters that are more likely to lead to safety violations. The epigenetic model can achieve this due to several reasons. 
First, to better understand the environment, the epigenetic model integrates a multi-modal fusion transformer that leverages a self-attention mechanism to integrate geometric and semantic information from the environment across multiple modalities. This step extracts a high-level feature representation of the environment for the GS probability generation. 
Second, the self-attention layer for the GS probability generation (Section~\ref{subsubsec:aeg}) allows the epigenetic model to weigh the importance of each environmental parameter in a given environment by calculating attention weights between the word embeddings of the environmental parameters and feature representations of the environmental states. The calculated attention weights determine the contribution of each environmental parameter to the safety violations, which is further used as GS probabilities.


\subsection{Incorporating multiple objectives}
The current \method design considers a single objective (i.e., distance minimization) when generating safety-critical AV test scenarios. In practice, multiple AV testing objectives related to safety and functionality often need to be considered simultaneously. As concluded in Section~\ref{subsec:result_rq1}, there are correlations between the objective and other metrics: for \co and \is, the correlation is very high, while for \rc and \ds the correlation is negligible. This observation inspires us to further investigate the multi-objective \method design to optimize multiple objectives simultaneously. 
Considering that some objectives are high-correlated (e.g., \md with \co and \is), one possibility could be deriving one objective by combining these objectives together, e.g., summing them up with equal weights. For others (e.g., \md with \rc and \ds), multi-objective search algorithms (MOSAs) such as NSGA-II~\cite{deb2002fast} can be applied to solve the optimization problem. However, no algorithms currently combine MOSAs and epigenetics, which calls for future research.

\subsection{Investigating other epigenetic mechanisms}
In this work, we employ gene silencing (i.e., GS) as the epigenetic mechanism (i.e., EM). In addition to GS, other EMs have also been studied in biology, such as Genomic Imprinting (GI)~\cite{reik2001genomic}, Paramutation~\cite{brink1973paramutation}, and X-Chromosome Inactivation (XCI)~\cite{lyon1999x}. Implementing these EMs as novel epigenetic operators and integrating them into GAs is interesting to investigate. For instance, GI involves parent-of-origin-specific gene expression patterns -- the activity of certain genes is influenced by whether they are inherited from the mother or the father. In the GA context, this could be analogous to assigning different levels of influence (reflected as weights or probabilities) to genes inherited from one parent during optimization, i.e., introducing the concept of "parental influence" in GAs, where certain parameters are more likely to be inherited from one parent than from the other. Integrating EMs, such as GI, into GAs opens up new possibilities for improving their performance, which we believe is interesting to investigate in the future.

\section{Related Work}\label{sec:related_work}
We discuss relevant works related to scenario-based AV testing (Section~\ref{sec:av_testing}) and epigenetic algorithms (Section~\ref{sec:epiga}). 

\subsection{Scenario-Based AV Testing}\label{sec:av_testing}
Various driving scenarios are needed to test how well an AV interacts with the environment and makes proper decisions. However, infinite driving scenarios make it impossible to exhaustively test AV. Thus, we must identify critical scenarios, i.e., scenario-based AV testing~\cite{10.1109/TSE.2022.3170122}. In the literature, a set of search-based testing (SBT) approaches have been proposed~\cite{ben2016testing,abdessalem2018testing,abdessalem2018ASE,calo2020generating,9251068} and also some reinforcement learning (RL) based AV testing approaches have been proposed as well~\cite{feng2023dense,9468363,lu2022learning,10.1109/ICSE48619.2023.00155}. There also exist works on generating scenarios from real-world driving data~\cite{zhang_risk_2022,yan2023learning,gambi2019generating}.

SBT-based approaches identify critical scenarios where fitness functions guide the optimization process toward generating critical scenarios. For example, NSGAII-SM~\cite{ben2016testing} combines NSGA-II~\cite{deb2002fast} with surrogate models to generate test scenarios for a pedestrian detection vision system.
NSGAII-DT~\cite{abdessalem2018testing} tests a vision-based control system combining NSGA-II with decision tree classification models to effectively identify critical scenarios.
Considering failures in AV may also originate from unintended interactions among system features (e.g., command conflicts between the automated emergency braking system and the adaptive cruise control system may compromise the AV's safety), 
Abdessalem et al.~\cite{abdessalem2018ASE} integrated a set of hybrid objectives with a search algorithm to generate critical scenarios.
Cal\`o et al.~\cite{calo2020generating} adapted NGSA-II to search for collisions and AV configurations that can avoid such collisions. 
By combining fuzzing testing and search algorithms, Li et al.~\cite{9251068} developed AV-FUZZER to generate critical scenarios that can identify safety violations of AV.
Focusing on testing AV against traffic laws, Sun et al.~\cite{10.1145/3551349.3556897} proposed LawBreaker, which adopts a fuzzing engine to search for scenarios that can effectively violate traffic laws.

RL-based testing approaches identify critical scenarios by dynamically and adaptively exploring the vast parameter space. 
Chen et al.~\cite{9468363} tests lane-changing models and developed an RL-based adaptive testing framework to generate time-sequential adversarial scenarios. DeepCollision~\cite{lu2022learning} is an RL-based approach that generates safety-critical scenarios by dynamically configuring an AV's operating environment. By combining RL and multi-objective search, Haq et al.~\cite{10.1109/ICSE48619.2023.00155} proposed a multi-objective RL approach, MORLOT, for testing AV. MORLOT uses RL to adaptively generate critical scenarios that can cause requirement violations and adopts multi-objective search to cover as many requirements as possible.
Feng et al.~\cite{feng2023dense} adapted dense RL, in which the Markov decision process is edited by removing non-safety-critical states, to learn critical scenarios from naturalistic driving data.

Approaches have also been proposed to identify critical scenarios from traffic accident reports and real-world driving data~\cite{zhang_risk_2022,yan2023learning}. Gambi et al.~\cite{gambi2019generating} proposed AC3R to extract the vehicle crash information from the police reports and recreate the scenario in simulation for testing AVs. Zhang et al.~\cite{zhang_risk_2022} adopted surrogate models in a scenario-based test to expedite the risk assessment of AV, and to fit the naturalistic distribution of the generated scenarios, HighD dataset~\cite{krajewski2018highd} was applied in their approach. Zhang et al.~\cite{10.1145/3597926.3598070} proposed the M-CPS model to extract information from real-world accidents. Based on the extracted accident information, a mutation testing solution automatically builds critical testing scenarios. To bridge the gap between the simulation and the real-world driving environment, Yan et al.~\cite{yan2023learning} proposed NeuralNDE, which learns naturalistic multi-agent interaction behavior from vehicle trajectory data based on statistical realism.

Different from the existing methods that treat environmental parameters equally when generating critical scenarios, \method utilizes a novel attention-based GS mechanism to selectively express parameters with high contribution to safety violations and silence those with low contribution; therefore, compared to the baseline methods (i.e., \ga and \methodequal), \method is more effective and efficient in generating critical scenario generation (Section~\ref{subsec:result_rq1} and Section~\ref{subsec:result_rq2}).

\subsection{Epigenetic Algorithm}\label{sec:epiga}
In the literature, only a few epigenetic algorithms have been proposed. 
Tanev and Yuta~\cite{tanev_epigenetic_2008} proposed an epigenetic programming approach, i.e., epigenetic learning (EL), incorporating histone modification mechanisms to control gene expression.
EpiGA~\cite{STOLFI2018250} implements GS and integrates it into GA to control how genes are expressed in response to environmental uncertainties. Another work, EpiLearn~\cite{mukhlish2020reward}, encodes dynamic environmental changes as an epigenetic layer in a learning process to allow for adaptive and efficient learning. RELEpi~\cite{mukhlish2020reward2} supports the coevolving decision-making of groups of agents (swarms) in uncertain environments, but it has not yet proven effective for solving real-world problems. These works demonstrate that epigenetic algorithms are a promising direction for coping with uncertainty, which has also been emphasized in~\cite{VERSE2023}.

Our approach \method is the first to incorporate epiGA and extend its GS mechanism for addressing AV testing challenges by adopting a transformer model to effectively identify each environmental parameter's contribution to safety violations as GS probabilities.

\section{Conclusion and Future Work}\label{sec:conclusion}
Given infinite driving scenarios, generating critical ones for scenario-based AV testing is practically challenging. In this paper, we propose a novel method, named \method, which extends the Gene Silencing (GS) mechanism in \epiga to selectively express or silence certain environmental parameters with GS probabilities. By using an epigenetic model built based on an attention mechanism, the GS probabilities are adaptively generated as the driving environment changes. We evaluate \method on a state-of-the-art AV using two comparison baselines, and the experiment results show that \method outperformed the baselines in terms of collision scenario generation while guaranteeing a faster convergence speed. 
In the future, we are interested in studying the generalization of \method by conducting experiments with other AVs and simulators. In addition, we plan to investigate other epigenetic mechanisms that can be potentially applied in \method, such as histone modifications and genomic imprinting.





\ifCLASSOPTIONcaptionsoff
  \newpage
\fi

\balance

\bibliographystyle{ieeetr}
\bibliography{references}

\end{document}

%% file: header.tex
\usepackage{xspace}
\usepackage{wrapfig}
\usepackage{lipsum}
\usepackage{multirow}
\usepackage{graphicx}
\usepackage{float} 
\usepackage{threeparttable}
\usepackage{makecell}
\usepackage{colortbl}
\usepackage{listings}
\usepackage{xcolor}
\usepackage{subcaption}
\usepackage{booktabs}
\usepackage{algorithm}
\usepackage{algpseudocode}
\usepackage{amsmath}
\usepackage{mathrsfs}  
\usepackage{enumitem}
\usepackage{bbding}
\usepackage{fdsymbol}
\usepackage{tikz}
\usepackage{url}
\usepackage{balance}
\usepackage{paralist}
\usepackage{threeparttable}
\usepackage{makecell}

\newcommand{\method}{\textit{EpiTESTER}\xspace}
\newcommand{\ads}{ADS\xspace}
\newcommand{\avut}{AVUT\xspace}
\newcommand{\npc}{NPC vehicle\xspace}
\newcommand{\epiga}{epiGA\xspace}
\newcommand{\methodequal}{$\textit{EpiTESTER}_{eq}$\xspace}

\newcommand{\ga}{\textit{GA}\xspace}

\newcommand{\md}{\textit{MD}\xspace}
\newcommand{\co}{\textit{CO}\xspace}
\newcommand{\rc}{\textit{RC}\xspace}
\newcommand{\is}{\textit{IS}\xspace}
\newcommand{\ds}{\textit{DS}\xspace}

%% file: algs/nbr.tex
\begin{algorithmic}[1]
\Function{NucleosomeBasedReproductionCell}{$c_1$, $c_2$}
  \State $x_1$, $x_2 \leftarrow \textsc{GetSolution}(c_1), \textsc{GetSolution}(c_2)$
  \State $n_1$, $n_2$ $\leftarrow$ $\textsc{getNucleosome}(c_1)$, $\textsc{getNucleosome}(c_2)$ 
  \State $n \leftarrow n_1 \textbf{OR} n_2$
  \For{$j$ $\in$ ($1, \textsc{Len}(n))$}
    \If{$n(j)$}
    \State $x_1^\prime(j)\leftarrow x_1(j)$
    \State $x_2^\prime(j)\leftarrow x_2(j)$
    \Else
    \State $x_1^\prime(j)\leftarrow x_2(j)$
    \State $x_2^\prime(j)\leftarrow x_1(j)$
    \EndIf
  \EndFor
  \State $c_1^\prime, c_2^\prime \leftarrow \textsc{Cell}(x1^\prime, n), \textsc{Cell}(x2^\prime, n)$
  \State \Return $c_1^\prime, c_2^\prime$
\EndFunction

\end{algorithmic}

%% file: algs/gs.tex
\begin{algorithmic}[1]
\Function{GeneSilencingCell}{$c$, $Pr_{gs}$}
  \State $x$ $\leftarrow \textsc{GetSolution}(c)$
  \State $n_1$ $\leftarrow$ $\textsc{getNucleosome}(c)$
  \For{$j$ $\in$ ($1, \textsc{Len}(n))$}
    \If{$n(j) \land rnd() < Pr_e$}
        \If{$rnd() > Pr_{gs}(j)$}
   \State $\textsc{Express}(x(j))$
        \EndIf
    \EndIf
  \EndFor
  \State \Return $c_1^\prime, c_2^\prime$
\EndFunction

\end{algorithmic}

%% file: figures/scenarios.tex
\begin{subfigure}[b]{0.25\textwidth}
    \centering
    \includegraphics[width=\textwidth]{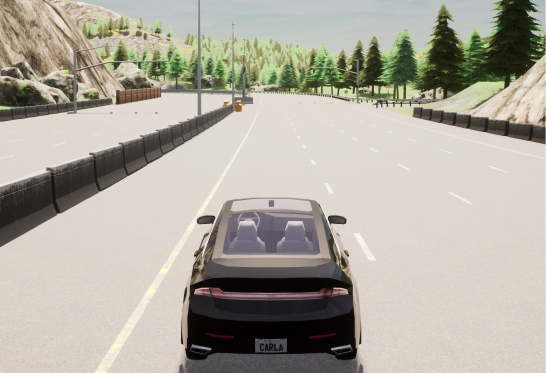}
    \caption{\centering {$Env_1$}}
    \label{fig:s1}
\end{subfigure}
\hfill
\begin{subfigure}[b]{0.25\textwidth}
    \centering
    \includegraphics[width=\textwidth]{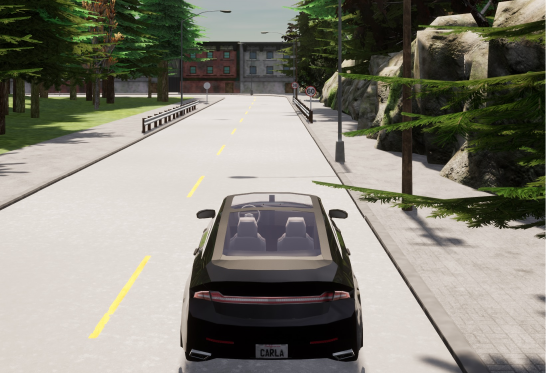}
    \caption{\centering {$Env_2$}}
    \label{fig:s2}
\end{subfigure}
    \hfill
\begin{subfigure}[b]{0.25\textwidth}
    \centering
    \includegraphics[width=\textwidth]{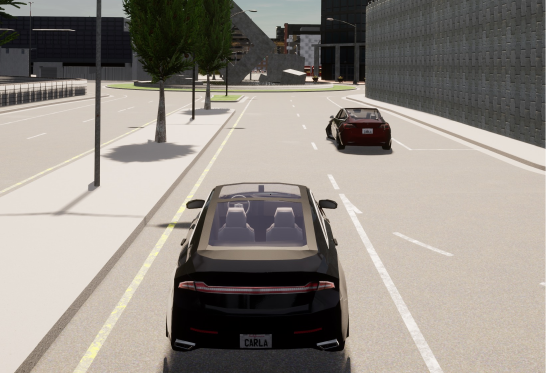}
    \caption{\centering {$Env_3$}}
    \label{fig:s3}
\end{subfigure}
     \hfill
\begin{subfigure}[b]{0.25\textwidth}
    \centering
    \includegraphics[width=\textwidth]{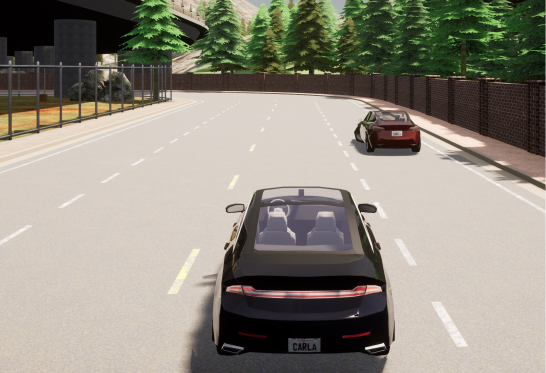}
    \caption{\centering {$Env_4$}}
    \label{fig:s4}
\end{subfigure}

%% file: tables/table1.tex
\begin{tabular}{cccccccccccc}
\toprule
\multirow{2}{*}{\begin{tabular}[l]{@{}l@{}}\textit{Initial} \\ \textit{Env.}\end{tabular}} & \multicolumn{1}{l}{\multirow{2}{*}{\begin{tabular}[l]{@{}l@{}}\method \textit{vs.}\\ \textit{baselines}\end{tabular}}} & \multicolumn{2}{c}{$MD\downarrow$}   & \multicolumn{2}{c}{$CO\uparrow$}   & \multicolumn{2}{c}{$RC\downarrow$}   & \multicolumn{2}{c}{$IS\downarrow$}   & \multicolumn{2}{c}{$DS\downarrow$}   \\ \cmidrule(r){3-4} \cmidrule(r){5-6}\cmidrule(r){7-8}\cmidrule(r){9-10}\cmidrule(r){11-12}
                                     & \multicolumn{1}{c}{}                        & $\hat{A}_{12}$    & \textit{p-value}               & $\hat{A}_{12}$    & \textit{p-value}               & $\hat{A}_{12}$    & \textit{p-value}               & $\hat{A}_{12}$    & \textit{p-value}               & $\hat{A}_{12}$    & \textit{p-value}               \\ \midrule
\multirow{2}{*}{$Env_1$}    & \ga &                                  \textbf{0.347} &  \textbf{$<$0.05} &  \textbf{0.658} &  \textbf{$<$0.05} &  0.551× &  $<$0.05 &   \textbf{0.388} &  \textbf{$<$0.05} &   0.522 &  0.337 \\
& \methodequal &   \textbf{0.468} &  \textbf{$<$0.05} &  \textbf{0.535} &  \textbf{$<$0.05} &   \textbf{0.449} &  \textbf{$<$0.05} &   \textbf{0.465} &  \textbf{$<$0.05} &   \textbf{0.445} &  \textbf{$<$0.05} \\ \midrule
\multirow{2}{*}{$Env_2$}    & \ga &                                  \textbf{0.302} &  \textbf{$<$0.05} &  \textbf{0.698} &  \textbf{$<$0.05} &  0.564× &  $<$0.05 &   \textbf{0.302} &  \textbf{$<$0.05} &   0.481 &  0.413 \\
& \methodequal &   \textbf{0.418} &  \textbf{$<$0.05} &  \textbf{0.573} &  \textbf{$<$0.05} &  0.578× &  $<$0.05 &   \textbf{0.427} &  \textbf{$<$0.05} &   0.540 &  0.089 \\ \midrule
\multirow{2}{*}{$Env_3$}    & \ga &                                  \textbf{0.387} &  \textbf{$<$0.05} &  \textbf{0.613} &  \textbf{$<$0.05} &  0.608× &  $<$0.05 &   \textbf{0.454} &  \textbf{$<$0.05} &   0.458 &  0.068 \\
& \methodequal &   0.502 &  0.699 &   0.498 &  0.705 &   \textbf{0.455} &  \textbf{$<$0.05} &   0.502 &  0.705 &   0.459 &  0.065 \\ \midrule
\multirow{2}{*}{$Env_4$}    & \ga &                                  \textbf{0.339} &  \textbf{$<$0.05} &  \textbf{0.662} &  \textbf{$<$0.05} &   \textbf{0.268} &  \textbf{$<$0.05} &   0.477 &  0.231 &   \textbf{0.240} &  \textbf{$<$0.05} \\
& \methodequal &   0.497 &  0.467 &   0.503 &  0.478 &   \textbf{0.291} &  \textbf{$<$0.05} &  0.546× &  $<$0.05 &   \textbf{0.292} &  \textbf{$<$0.05} \\ \midrule \midrule

\multirow{2}{*}{$Env_{1-4}$}    & \ga &                                  \textbf{0.481} &  \textbf{$<$0.05} &  \textbf{0.519} &  \textbf{$<$0.05} &   \textbf{0.461} &  \textbf{$<$0.05} &   \textbf{0.479} &  \textbf{$<$0.05} &   \textbf{0.459} &  \textbf{$<$0.05} \\
& \methodequal &   \textbf{0.351} &  \textbf{$<$0.05} &   \textbf{0.650} &  \textbf{$<$0.05} &   0.480 &  0.091 &   \textbf{0.391} &  \textbf{$<$0.05} &   \textbf{0.433} &  \textbf{$<$0.05} \\ \bottomrule
\end{tabular}

%% file: tables/cor_test.tex
\begin{tabular}{ccllll}
\toprule
\multirow{2}{*}{\begin{tabular}[l]{@{}l@{}}\textit{Initial} \\ \textit{Env.}\end{tabular}}    & \multirow{2}{*}{\textit{Method}}  &  \multicolumn{4}{c}{\textit{MD$\downarrow$ vs.}} \\ \cmidrule(r){3-6}
   &      &  \textit{CO$\uparrow$} &  \textit{RC$\downarrow$} &  \textit{IS$\downarrow$} &  \textit{DS$\downarrow$} \\ \midrule
\multirow{3}{*}{$Env_1$} &    \method &  -0.999 $\begingroup\color{blue}\blacktriangle\endgroup$ &  -0.129 &   0.999 $\begingroup\color{blue}\blacktriangle\endgroup$ &   0.051 $\bullet$ \\
                              &   \ga &  -0.963 $\begingroup\color{blue}\blacktriangle\endgroup$ &  -0.574 &   0.904 $\begingroup\color{blue}\blacktriangle\endgroup$ &  -0.484 \\
                       & \methodequal &  -0.996 $\begingroup\color{blue}\blacktriangle\endgroup$ &   0.282 &   0.996 $\begingroup\color{blue}\blacktriangle\endgroup$ &   0.424 \\ \midrule
\multirow{3}{*}{$Env_2$} &    \method &  -0.983 $\begingroup\color{blue}\blacktriangle\endgroup$ &  -0.698 &   0.983 $\begingroup\color{blue}\blacktriangle\endgroup$ &  -0.134 \\
                              &   \ga &  -0.829 $\begingroup\color{blue}\blacktriangle\endgroup$ &  -0.523 &   0.829 $\begingroup\color{blue}\blacktriangle\endgroup$ &  -0.202 \\    
                      &  \methodequal &  -0.953 $\begingroup\color{blue}\blacktriangle\endgroup$ &  -0.764 $\begingroup\color{blue}\blacktriangle\endgroup$ &   0.953 $\begingroup\color{blue}\blacktriangle\endgroup$ &  -0.312 \\ \midrule
\multirow{3}{*}{$Env_3$} &    \method &  -1.000 $\begingroup\color{blue}\blacktriangle\endgroup$ &  -0.189 &   1.000 $\begingroup\color{blue}\blacktriangle\endgroup$ &   0.209 \\
                              &   \ga &  -0.988 $\begingroup\color{blue}\blacktriangle\endgroup$ &  -0.572 &   0.848 $\begingroup\color{blue}\blacktriangle\endgroup$ &   0.546 \\     
                       & \methodequal &  -1.000 $\begingroup\color{blue}\blacktriangle\endgroup$ &  -0.180 &   1.000 $\begingroup\color{blue}\blacktriangle\endgroup$ &   0.186 \\ \midrule
\multirow{3}{*}{$Env_4$} &    \method &  -0.999 $\begingroup\color{blue}\blacktriangle\endgroup$ &   0.138 &   0.999 $\begingroup\color{blue}\blacktriangle\endgroup$ &   0.392 \\
                              &   \ga & -0.973 $\begingroup\color{blue}\blacktriangle\endgroup$ &  -0.392 &   0.845 $\begingroup\color{blue}\blacktriangle\endgroup$ &  -0.262 \\
                       & \methodequal &  -1.000 $\begingroup\color{blue}\blacktriangle\endgroup$ &  -0.136 &   0.398 &   0.033 $\bullet$ \\ 
                       
                       \midrule \midrule
\multirow{3}{*}{$Env_{1-4}$} &   \method &  -0.998 $\begingroup\color{blue}\blacktriangle\endgroup$ &  -0.209 &   0.673 &   0.118 \\
                              &   \ga &     -0.957 $\begingroup\color{blue}\blacktriangle\endgroup$ &  -0.442 &   0.874 $\begingroup\color{blue}\blacktriangle\endgroup$ &  -0.107 \\
                       & \methodequal &     -0.996 $\begingroup\color{blue}\blacktriangle\endgroup$ &  -0.304 &   0.743 $\begingroup\color{blue}\blacktriangle\endgroup$ &   0.021 $\bullet$ \\ \bottomrule
\end{tabular}